\documentclass[galaxies,article,accept,moreauthors,pdftex,10pt,a4paper,natbib]{mdpi} 
\pdfoutput=1 
\firstpage{1} 
\articlenumber{6}
\doinum{10.3390/galaxies6040130}
\pubvolume{130}
\pubyear{2018}
\copyrightyear{2018}
\history{Received: 31 October 2018; Accepted: 27 November 2018; Published: 30 November 2018}
\pdfoutput=1

\usepackage{comment}
\usepackage{multirow}
\usepackage{subcaption}

\usepackage[T1]{fontenc}
\usepackage[latin9]{luainputenc}
\usepackage{array}

 \theoremstyle{mdpi}
 \newcounter{thm}
 \newcounter{ex}
 \newcounter{re}
 \theoremstyle{mdpidefinition}

\def\gtsima{$\; \buildrel > \over \sim \;$}
\def\gsim{\lower.5ex\hbox{\gtsima}}

\def\lsima{$\; \buildrel < \over \sim \;$}
\def\lsim{\lower.5ex\hbox{\lsima}}

\Title{The Rate of Short-Duration Gamma-Ray Bursts in the Local Universe}

\Author{Soheb Mandhai $^{1,\dagger,}$*, Nial Tanvir $^{1,\dagger}$, Gavin Lamb $^{1,\dagger}$, Andrew Levan $^{2}$ and David Tsang $^{3}$}

\address{%
$^{1}$ \quad Department of Physics and Astronomy, University of Leicester, University Road, LE1 7RH, U.K.; nrt3@leicester.ac.uk (N.T.); \mbox{gpl6@leicester.ac.uk (G.L.)}\\
$^{2}$ \quad Department of Physics, University of Warwick, Coventry, CV4 7AL, U.K.; A.J.Levan@warwick.ac.uk\\
$^{3}$ \quad Department of Physics, University of Bath, Claverton Down, Bath, BA2 7AY, U.K.; D.Tsang@bath.ac.uk
}

\corres{Correspondence: sfm13@leicester.ac.uk; nrt3@leiceseter.ac.uk}

\firstnote{These authors contributed equally to this work.}

\abstract{
Following the faint gamma-ray burst, GRB\,170817A, coincident with a gravitational wave-detected binary neutron star merger at $d\sim40$\,Mpc, we consider the constraints on a local population of faint short duration GRBs (defined here broadly as $T_{90}<4$\,s).
We review proposed low-redshift short-GRBs and consider statistical limits on a $d\lsim200$\,Mpc population using {Swift}/Burst Alert Telescope (BAT), {Fermi}/Gamma-ray Burst Monitor (GBM)
, and Compton Gamma-Ray Observatory (CGRO) Burst and Transient Source Experiment (BATSE) GRBs.
{Swift}/BAT short-GRBs give an upper limit for the all-sky rate of $<4$\,y$^{-1}$ at $d<200$\,Mpc, corresponding to $<5$\% of SGRBs.
Cross-correlation of selected {CGRO}/BATSE and {Fermi}/GBM GRBs with $d<100$\,Mpc galaxy positions
returns a weaker constraint of
$\lsim12\,{\rm y^{-1}}$.
A separate search for correlations due to SGR giant flares in nearby ($d<11$\,Mpc) galaxies finds an upper limit of $<3\,{\rm y^{-1}}$.
Our analysis suggests that GRB\,170817A-like events are likely to be rare in existing SGRB catalogues.
The best candidate for an analogue remains GRB\,050906, where the {Swift}/BAT location was consistent with the galaxy IC\,0327 at $d\approx132$\,Mpc.
If binary neutron star merger rates are at the high end of current estimates, then our results imply that at most a few percent will be accompanied by detectable gamma-ray flashes in the forthcoming LIGO/Virgo science runs.
}

\begin{document}

%%%%%%%%%%%%%%%%%%%%%%%%%%%%%%%%%%%%%%%%%%

%%%%%%%%%%%%%%%%%%%%%%%%%%%%%%%%%%%%%%%%%%

\section{Introduction}\label{s:intro}

Gamma-ray bursts are classified as either long or short duration.
The distinction is most clearly indicated by the time taken to receive $90\%$ of the total gamma-ray fluence, the $T_{90}$ of the burst.
The canonical definition places bursts having $T_{90}>2$\,s in the long-duration class and those with $T_{90}<2$\,s in the short-duration (SGRB) class \citep{kouveliotou1993}, although in practice, there is an overlap in the population properties, and measured duration is also detector dependent (e.g.,~\citep[]{Bromberg2013}).

The short class is thought to arise predominantly during the gravitational wave-driven merger of compact binary objects, particularly binary neutron stars (NSNS) or binary stellar-mass black hole and neutron star systems (NSBH) \citep{eichler1989,narayan1992,mochkovitch1993,bogomazov2007,roberts2011electromagnetic,giacomazzo2012compact,paschalidis2017general}. 
The rapid accretion of material disrupted during the merger will launch an ultra-relativistic jet, which gives rise to an SGRB for observers aligned within the opening angle of the jet.
The compact binary merger scenario is supported by the broad population of SGRB host galaxies, both with and without recent star formation \citep{berger2014short}, and the wide range of offsets between burst location and host \citep{fong2013locations,tunnicliffe2013nature}. 
These host offsets suggest that there is a delay between formation and merger and that such systems can sometimes receive large natal kicks from the supernovae that form the compact objects \citep{bray2016}.
In a handful of cases, possible transients powered by nucleosynthesis in the neutron-rich ejecta---so called ``kilonovae'' (often called ``macronovae'')---have also been detected (e.g.,~\citep[]{tanvir2013, berger2013r, jin2015light, gompertz2018}). 

The detection of the short-duration GRB\,170817A accompanying the gravitational wave-detected merger GW170817 of a binary neutron star system consolidated the idea that compact binary mergers are the progenitors of SGRBs (e.g., \citep[]{abbott2017}).
However, GRB\,170817A was unusual when compared to previous short-bursts for which the redshift has been determined.
In particular, at $d\approx40$\,Mpc, GRB\,170817A was much closer and intrinsically less luminous than any previous SGRB with a securely-measured distance.
Nonetheless, this discovery has revived interest in the rate of SGRBs (and events that are phenomenologically similar) in the local universe ($d\lsim200$\,Mpc), which has particular bearing on the expected fraction of gravitational wave detections that will be accompanied by detectable gamma-ray flashes \citep[e.g.][]{beniamini2018b}.

Previous work, looking at the spatial correlation between galaxies in the local universe and the positions in the sky of SGRBs from the Compton Gamma-Ray Observatory (CGRO) Burst and Transient Source Experiment (BATSE) catalogue concluded that as many as 10--25\% of $T_{90}<2$\,s bursts could be of a local origin \citep{tanvir2005origin,chapman2009short}.
However, this correlation result was of comparatively weak statistical significance, while the poor positional information (a few degrees of precision) provided by BATSE was not sufficient to identify any given burst with a particular host galaxy.
A rate of binary mergers as high as 25\% would have significant implications for the expected detection rates in future GW observations and also for their contribution to the heavy element nucleosynthesis budget. As such, it is important to re-assess the local SGRB rate in light of more recent observational evidence.
Since 2005, the {Swift} satellite has localised $\gsim100$ SGRBs with at least a few arcmin astrometric precision, with none being clearly associated with a local galaxy. This certainly suggests that the rate of local SGRBs must be below that estimated from the original cross-correlation analysis.
In this paper, we revisit this question,
first discussing a range of potential low-luminosity SGRB-like events, then
considering the observational constraints on such populations based on the samples of bursts seen by {Swift}, {CGRO}/BATSE,
{Fermi}/GBM
, and the Inter-Planetary Network.

\section{Potential Low-Redshift Short Gamma-Ray Transients}\label{s:progenitors}

\subsection{Short Gamma-Ray Bursts}

The observed population of SGRBs with measured redshift spans a range from $z\sim0.1$ (e.g.,~\citep[]{rowlinson2010unusual}) to $z>1$ (e.g.,~\citep[]{Gorosabel2006,selsing2018host}), based primarily on events that have been well-localised by the {Neil Gehrels Swift Observatory} in nearly 14 years of observation to date.
Thus, it seems that the rate of on-axis SGRBs with detectable gamma-ray emission in the local universe ($z\lsim0.05$) is likely to be low. However, the~faint-end of the cosmological SGRB luminosity function is not well constrained, and it may be that faint SGRBs without a redshift measurement contribute to a local population if they are viewed off-axis slightly outside of the jet core \citep{lamb2017, jin2018, kathirgamaraju2018}.

\subsection{GW170817-Like Events}

It is now clear, following GRB\,170817A, that some NSNS mergers can also produce weak events.
Since there is good evidence that our sight-line was well off
the primary jet axis for GW170817 \citep{mooley2018, troja2018}
, it is more likely that the observed burst of gamma-rays in this event
was produced by a shock-breakout of a cocoon, rather than being emission from internal shocks in the jet (e.g.,~\citep[]{lazzati2017, kasliwal2017,gottlieb2018,lamb2018}).

GRB\,170817A has re-ignited interest in the local rate of SGRBs $< 200\ \textrm{Mpc}$, 
approximately the sensitivity limit to NSNS mergers for the current generation of gravitational wave detectors.
GRB 170817A-like transients
could produce a population of faint SGRBs with durations of a few seconds.
Bursts with a comparable energy to GRB 170817A could be detectable in gamma-rays at distances of $\sim 100$ Mpc with current technology,
which is well matched to that of the current generation of gravitational wave interferometers.

\subsection{NSNS Merger Precursors}

A small fraction $\lsim$10\% of SGRBs seem to be preceded by a detectable precursor (a short and faint flash of gamma-rays) \citep{troja2010}.
These precursors occur 1--10\,s before the SGRB at a time when the pre-merger NSs are strongly interacting.
Tidal deformation of the merging NS crusts that exceeds the breaking strain will create cracks that can result in the isotropic emission of gamma-rays at a comparable energy to the observed precursors \citep{kochanek1992}.
Alternatively, this tidal mechanism may shatter the crust due to the excitation of a resonant mode during the periodic deformation, a resonant shattering flare (RSF) \citep{tsang2013}.

Alternative explanations for burst precursors include the breakout of a shock-wave produced by the NSNS collision \citep{kyutoku2014} or a pair fireball created by magnetospheric interaction between the merging NSs \citep{metzger2016}.
A burst of gamma-rays produced by these precursor mechanisms would be emitted isotropically and will result in a faint and local population of SGRB-like transients.
A population of faint and short $\leq0.5$-s gamma-ray transients could be apparent with a similar host-galaxy type association and offset as the general SGRB population.

\subsection{Giant Flares from Soft Gamma-Ray Repeaters}

An entirely different class of event is expected to contribute to a local population of faint SGRB-like transients,
namely giant flares (GF) from soft-gamma-ray repeaters (SGRs). 
For the purposes of this work, we are interested in the rate of SGR GFs in external galaxies as a potential contaminant of the SGRB catalogues that we are considering.

Highly magnetised neutron stars, or magnetars, were confirmed as the origin for SGRs with the observation of several outbursts from SGR 1900+14 after a prolonged period of quiescence \citep{kouveliotou1999}.
On~extremely rare occasions, an SGR will emit a giant flare with energy $\sim$1000-times greater than a regular SGR outburst.

On 27 December 2004, such a giant flare (GF) erupted from the magnetar SGR 1806-20 \citep{palmer2005, hurley2005}.
Somewhat less powerful giant flares had previously been observed from both SGR $0526-66$ and SGR $1900+14$ \citep{hurley2011}.
The isotropic equivalent energy of the SGR 1806-20 flare was initially reported as $2\times10^{46}$ erg, based on an estimated distance of 15\,kpc, meaning that similar events could potentially be observable by BATSE to a distance of $\sim30-40$ Mpc \citep{palmer2005, hurley2005}.
This raised speculation that a proportion of detected SGRBs might in fact be SGR GFs in low redshift galaxies.
The distance estimate for SGR 1806-20 was later revised by Bibby {et al.} \cite{bibby2008} to 8.7 kpc, reducing the peak luminosity estimate by a factor of $\sim$3 and the maximum distance that such a flare would be observable to $\sim$20--25 Mpc.

With only three detections of GFs in the MW and LMC over a period of 40--50 years, SGR GFs must be reasonably rare events.
Of the known SGR GFs in the Milky Way (MW) and Large Magellanic Cloud (LMC), the durations are in the range 0.2--0.5 s \citep{cline1980, thompson2001, hurley2005}.
Although this time scale is not directly comparable to the BATSE $T_{90}$ durations, it does suggest that SGR GFs should typically be $<1$ s in duration.

\section{ SGRBs Observed by {Swift}}

\begin{figure}[H]
\centering
\includegraphics[scale=0.6]{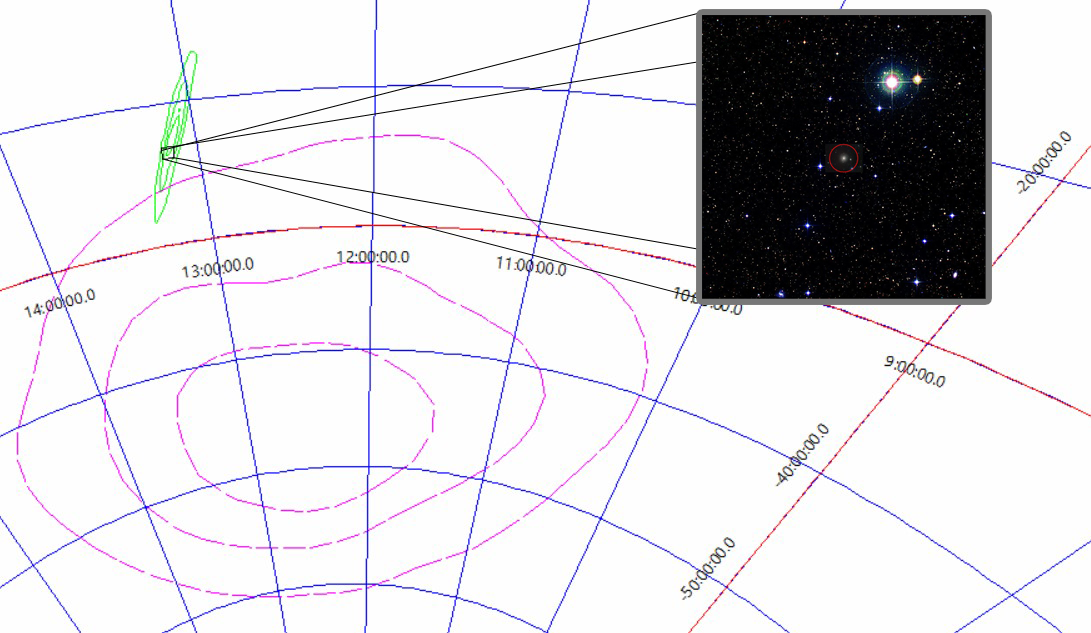}
\caption{The GRB\,170817A localisation by Fermi (magenta contours) in comparison to a typical {Swift}-BAT instrument localisation (red circle in zoomed panel) if the burst had been within the BAT field-of-view. The BAT region corresponds to a localisation with a radius of 3', and the inset panel is based on VLT
/VIMOS imaging reported in \citet{tanvir2017}. The LIGO/Virgo localisation for GW170817 is shown as green contours, with a one-square degree box centred on the origin of the event, the galaxy NGC\,4993.}
\label{fig:loc-comp}
\end{figure}

The {Neil Gehrels Swift Observatory} ({Swift}) is dedicated to detecting and following up on gamma-ray burst events. The on-board instruments consist of the Burst Alert Telescope (BAT), the X-ray Telescope (XRT), and the UV/Optical Telescope (UVOT) \citep{gehrels2004swift}. The BAT instrument observes $\sim$100 GRBs per year, providing a positional accuracy of a few arcminutes within a (fully-coded) field-of-view of $\sim$1.4 \,sr. A burst trigger is usually followed by an automated re-pointing to bring the event location to within the fields of the narrow-field instruments, with typical slew times of 20--70\,s \citep{barthelmy2005burst}. The XRT and UVOT are then able to make high resolution follow-up observations of the GRB afterglow and reduce the positional uncertainty to an accuracy within 0.5--5 arcseconds \citep{zhang2006physical,brown2009ultraviolet}.
These capabilities mean that {Swift} generally can localise GRBs much more precisely than other missions.
To underline this point, if~the host galaxy NGC\,4993 for GRB\,170817A had been in the {Swift} Burst Alert Telescope (BAT) field of view when GW170817 occurred, then it would have been immediately identified as the likely host based on the low probability ($\approx0.03\%$) of finding such a bright galaxy by chance within a given BAT gamma-ray localisation circle of a few arcmin radius.
A scale comparison of a {Fermi} GBM error region for GRB\,170817A relative to a typical {Swift} localisation is shown in Figure \ref{fig:loc-comp}.

To date, there have been no {Swift} SGRBs that have been unambiguously associated with $d<200$\,Mpc host galaxies.
However, possible associations have been pointed out for
GRBs 050906~\citep{Levan08GRB050906}, 070809 \citep{070809GCN}, 090417A \citep{090417GCN}, and 111020A \citep{tunnicliffe2013nature} at distances $d\lsim400$\,Mpc (see Table \ref{tab:impact}).

In order to conduct a more systematic
survey, we have searched in
the 2MASS
 Redshift Survey (2MRS) catalogue for further potential low redshift hosts. We used 2MRS as it provides a uniform coverage of galaxies over $\approx$91\% of the sky and yields a 97.6\% redshift completeness to a limiting K-band magnitude of $K=11.75$ \citep{Huchra2012}.
For the burst sample, given that an off-axis GRB is likely to have a longer duration than a canonical SGRB and that GRB\,170817A was not produced by an on-axis jet and lasted $\sim$2\,s, we consider bursts with emission duration T$_{90}< 4\ \textrm{s}$, resulting in a sample of 157 events. In addition to this search, we also performed a visual examination of DSS-II
 (red) images of the regions around bursts, since some galaxies within this volume may be fainter than the 2MRS catalogue limit. We found no compelling examples beyond those already in the literature or found by the automated search.

For each burst, we found the 2MRS galaxy with the lowest impact parameter (projected distance on the plane of the sky) and within a distance of 5--200 Mpc,
using galaxy distances from HyperLEDA
~\citep{HyperLEDAIII} or the NASA/IPAC Extragalactic Database (NED) \citep{helou1991nasa}\footnote{We place a lower limit on the distance of 5 Mpc because the angular scales associated with galaxies closer than this suggest that a significant fraction of the sky is within 200\,kpc (in projection) of a galaxy within this distance horizon.
}.
We set an upper limit for the impact parameter of 200\,kpc, which allows for the possibility of
binaries that received
moderate natal kick velocities of $\sim$100\textrm{ km s}$^{-1}$ and large binary merger times $\gsim$ $10^9 \textrm{ years}$.
Note that the impact parameter calculation is less certain for cases where there was only a BAT localisation.
Finally, we removed any matches for which a more distant origin was already established from a close proximity to a higher redshift host.

\begin{table}[H]
\caption{List of Swift catalogued SGRB detections that have been paired with the closest 2MASS
 Redshift Survey (2MRS) galaxy. The galactic distances used have been obtained from the associated references. Further bursts for which tentative host galaxies have been suggested in the literature are listed below the table break.} 
\label{tab:impact}
\scalebox{0.87}[0.87]{

{\begin{tabular}{>{\centering}m{1.2cm}>{\centering}m{0.7cm}>{\centering}m{1.5cm}>{\centering}m{2.4cm}>{\centering}m{1cm}>{\centering}m{1.7cm}>{\centering}m{1cm}>{\centering}m{1.1cm}>{\centering}m{1.6cm}>{\centering}m{1.4cm}}
\toprule
\textbf{GRB} & \boldmath{$T_{90}$} \textbf{{(}s{)}} &\textbf{ Angular Separation {(}arcmin{)}} & \textbf{Closest Galaxy} & \textbf{Galaxy Type} & \textbf{Optical Bands (B/R
) {(}mag
{)} }& \textbf{J-Band {(}mag{)}} & \boldmath{$d$} \textbf{{(}Mpc{)}} &\textbf{ Impact Parameter {(}kpc{)}} & \boldmath{$E_{\rm iso}$} \textbf{\mbox{{(}$10^{46}$ ergs{)}}}\tabularnewline
\midrule
050906 & 0.26 & 2.0 & IC 0328 & Sc & 14.0 {(}B{)} & 12.2 & 132 \citep{HyperLEDAIII} & 77 $\pm$ 109 & 1.9\tabularnewline
100213A & 2.40 & 5.4 & PGC 3087784 & S0-a & 14.7 {(}B{)} & 11.3 & 78 \citep{HyperLEDAIII} & 123 & 39.9\tabularnewline
111210A & 2.52 & 6.0 & NGC 4671 & E & 13.4 {(}B{)} & 10.1 & 43 \citep{HyperLEDAIII} & 76 & 7.5\tabularnewline
120403A & 1.25 & 4.9 & PGC 010703 & Sc & 14.4 {(}B{)} & 12.1 & 133 \citep{HyperLEDAIII} & 192 $\pm$ 90
 & 38.2\tabularnewline
130515A & 0.29 & 8.5 & PGC 420380 & S0-a & 16.0 {(}B{)} & 12.3 & 73 \citep{Cosmicflows3} & 180 & 28.4\tabularnewline
160801A & 2.85 & 6.7 & PGC 050620 & Sa & 15.2 {(}B{)} & 12.4 & 59\citep{HyperLEDAIII} & 115 & 10.7\tabularnewline
\midrule
070809 & 1.30 & 2.0 & PGC 3082279 \citep{070809GCN} & Sa & 16.3 {(}B{)} & 13.5 & 
180 \citep{helou1991nasa} & 105 & 64.4\tabularnewline
090417A & 0.07 & 4.4 & PGC 1022875 \citep{090417GCN} & S0-a & 15.9 {(}B{)} & 13.4 & 
360 \citep{helou1991nasa} & 461 $\pm$ 292 & 24.5\tabularnewline
111020A & 0.40 & 2.3 & FAIRALL 1160 & Sa & $\sim14$ {(}R{)} & 11.7 & 
81 \citep{tunnicliffe2013nature} & 54 & 9.4\tabularnewline
\bottomrule
\end{tabular}}}
\end{table}

\begin{figure}[H]
\centering
\vbox{
\hbox{
\centering

\begin{subfigure}[h]{0.33\textwidth}

\includegraphics[width=\textwidth,height=5cm]{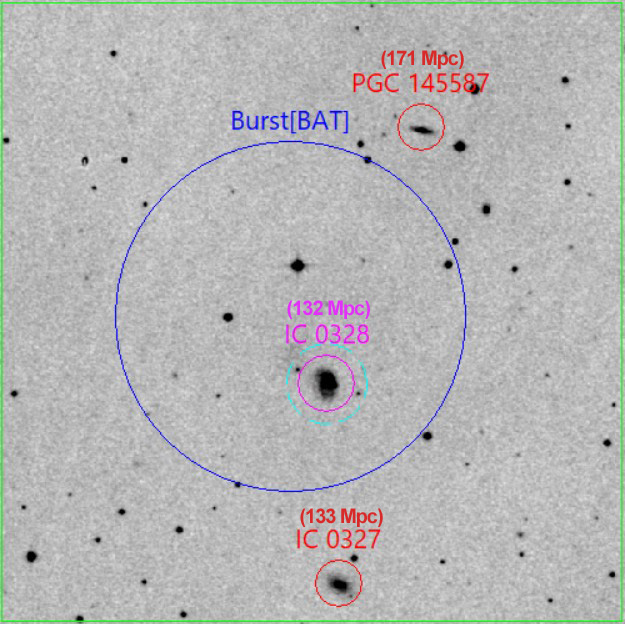}
\caption{GRB 050906$^{+}$}
\end{subfigure}
\begin{subfigure}[h]{0.32\textwidth}
\includegraphics[width=\textwidth,height=5cm]{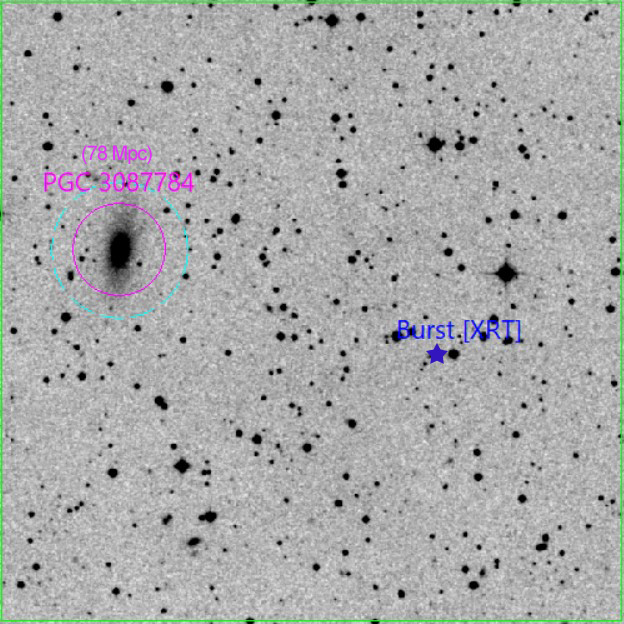}
\caption{GRB 100213A}
\end{subfigure}
\begin{subfigure}[h]{0.33\textwidth}
\includegraphics[width=\textwidth,height=5cm]{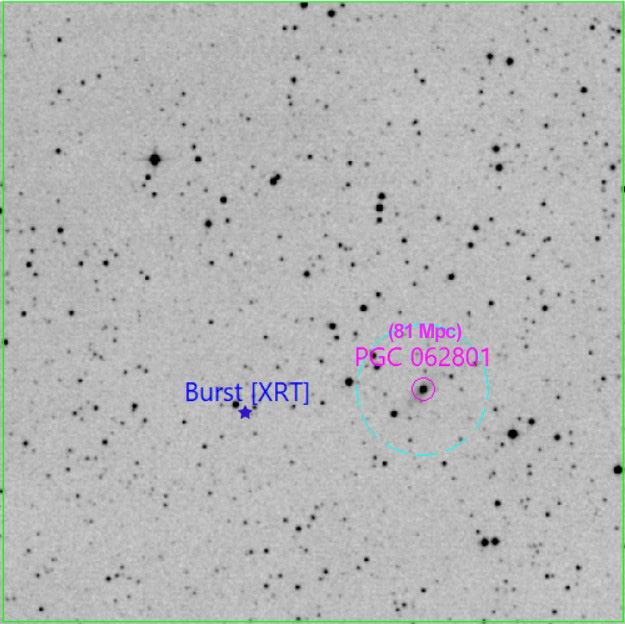}
\caption{GRB 111020A}
\end{subfigure}
\centering

}

\hbox{
\begin{subfigure}[h]{0.33\textwidth}
\includegraphics[width=\textwidth,height=5cm]{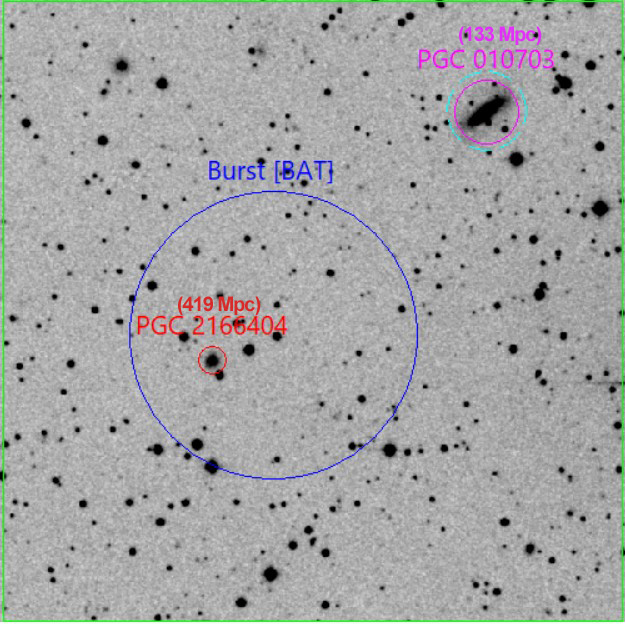}
\caption{GRB 120403A$^{+}$ }
\end{subfigure}
\begin{subfigure}[h]{0.32\textwidth}
\includegraphics[width=\textwidth,height=5cm]{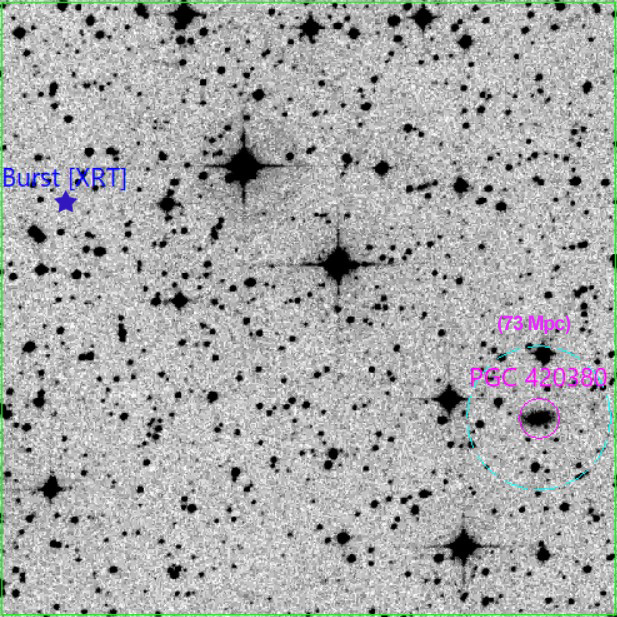}
\caption{GRB 130515A}
\end{subfigure}
\begin{subfigure}[h]{0.33\textwidth}
\includegraphics[width=\textwidth,height=5cm]{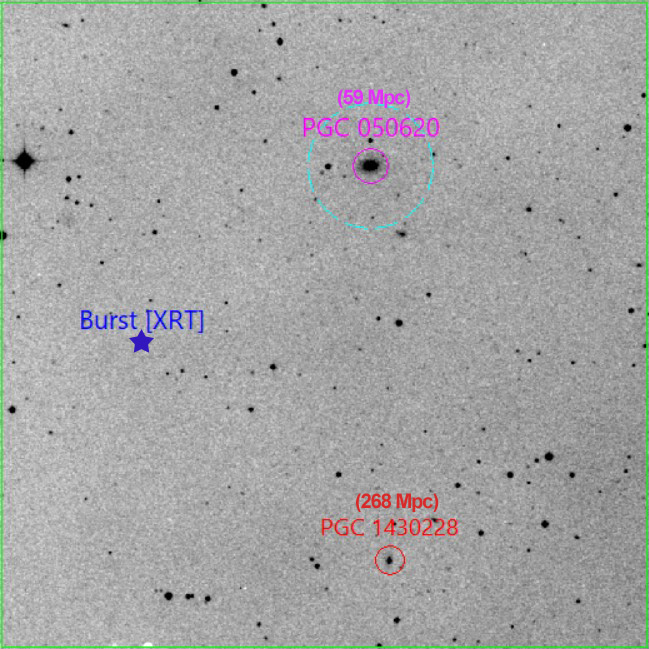}
\caption{GRB 160801A*}
\end{subfigure}

}

\hbox{
\begin{subfigure}[h]{0.33\textwidth}
\includegraphics[width=\textwidth,height=5cm]{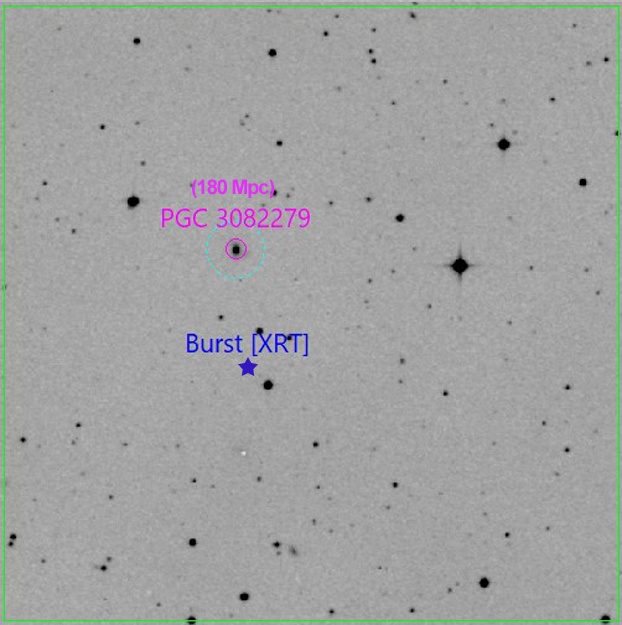}
\caption{GRB 070809}
\end{subfigure}

\begin{subfigure}[h]{0.32\textwidth}
\includegraphics[width=\textwidth,height=5cm]{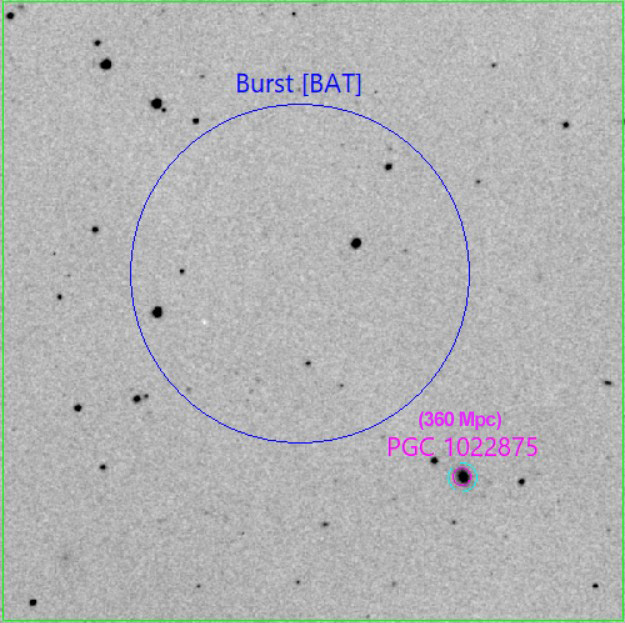}
\caption{GRB 090417A$^{+}$}
\end{subfigure}

\begin{subfigure}[h]{0.33\textwidth}
\includegraphics[width=\textwidth,height=5cm]{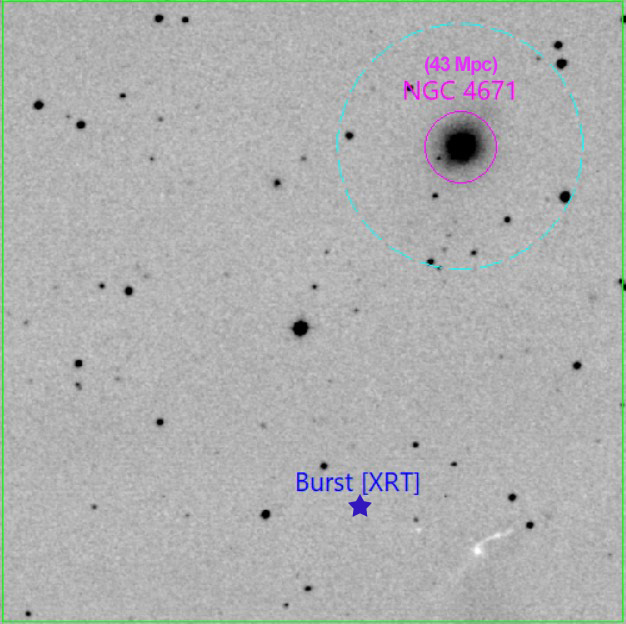}
\caption{GRB 111210A}
\end{subfigure}
}
}
\caption{Positions of {Swift}-detected GRBs with $T_{90}<4$\,s (blue) (from Table \ref{tab:impact}) plotted on DSS2 images of the fields, with the best candidate low redshift galactic host circled in magenta. Other potential candidates at greater distances are circled in red. A projected kick distance radius of 25 kpc is highlighted by the dashed cyan circle around the favoured host candidate. The standard panel size used for each source corresponds to an on-sky area of 10'\,$\times$\,10'.
* Corresponds to a 15' $\times$ 15' area.
$^{+}$ For events where an XRT localisation could not be obtained, either due to a delayed/absent slew (GRBs\,090417A, 120403A) or a lack of an X-ray afterglow (GRB\,050906), the Burst Alert Telescope (BAT) localisation is used.
}
\label{fig:SGRBs}
\end{figure}

This procedure finds potential nearby ($d<200$ Mpc) hosts for GRBs\,050906, 100213A, 111210A, 120403A, 130515A, and 160801A (full details are reported in Table~\ref{tab:impact} and shown in Figure \ref{fig:SGRBs}).
Of these, only GRB\,050906 had previously been noted in the literature\footnote{The candidate host for GRB\,111210A, Fairall 1160, is brighter than the 2MRS magnitude limit, but was erroneously classified as a star in the 2MASS database, and hence not included in 2MRS.}. It is interesting to note that if this host association is correct, then its isotropic energy was comparable to that of GRB\,170817A.
This case is also the only one for which the host candidate is within the positional uncertainty of the burst, implying that all the other cases would require the progenitor binaries to be kicked well outside their parent galaxies if the associations were real.
A high proportion of mergers occurring at large galactocentric radii would seem unlikely, for example being inconsistent with the comparatively low rate ($\sim$25\%) of hostless bursts found for SGRBs in the sample studied by \citet{Fong2013}.
Furthermore, repeating the experiment for large numbers of random positions showed that similar apparent associations are expected to occur by chance for $\sim$12 bursts in a sample of the size of our {Swift} sample.
 This suggests that the majority, quite possibly all, of these candidate associations, both from our 2MRS analysis and those reported previously, are likely to be chance alignments. In this regard, we note that several of the well-localised bursts from Table~\ref{tab:impact} have plausible alternative higher redshift hosts suggested in the literature, albeit that kicks would still be required to place the mergers outside the bodies of these galaxies (e.g., GRBs\,070809 and 111020A in \citep{tunnicliffe2013nature}, GRB\,111210A in \citep{Tanvir2011GCN}, and GRB\,130515A in \citep{Levan2013GCN}). Furthermore, in several cases, deep follow-up imaging places strong constraints on the luminosities of any associated kilonovae if the low redshift association were correct (e.g., GRB\,050906 in~\citep{Levan08GRB050906}, GRB\,070809 in \citep{Perley2008GCN}, GRB\,111020A in~\citep{Fong2012GRB111020A}, GRB\,130515A in \citep{Cenko2013GCN}).

Over the current elapsed mission duration of {Swift}, it has made roughly two years of all-sky observations with BAT (based on nearly 14 years in orbit and an instantaneous field of view of 10--15\% of the sky). Thus, from the evidence from {Swift}, we conclude a limit to the all-sky rate of detectable SGRBs
of $<$4\ \textrm{y}$^{-1}$ for $d < 200 \ \textrm{Mpc}$, with the likely rate being significantly less.
This could only be underestimated if low redshift mergers are typically happening at very large galactocentric distances, requiring large average kicks and long merger times, or if there is a preference for very faint hosts.

\section{SGRBs Observed by {CGRO}/BATSE and {Fermi}/GBM}\label{s:correlation}

Although {Swift} provides excellent positional accuracy, there have only been $\sim$\, 150 $T_{90}<4$\,s SGRBs detected over 14 years. {CGRO}/BATSE and {Fermi}/GBM have observed much larger samples of SGRBs, but with poorer localization; however, we can perform statistical analysis on the larger sample to constrain the fraction of the population that could arise from nearby galaxies.

The BATSE instrument on the space-based gamma-ray {CGRO} satellite continuously observed the unocculted sky from low Earth orbit, giving a field of view of $\sim2{\pi}$\,sr.
During its nine-year lifetime, BATSE detected $\sim$500 SGRBs with $T_{90}<2$\,s.
However, the large location uncertainties ($1\sigma$ errors, typically several degrees) effectively prevented identification of the galactic hosts of these bursts. By correlating a sample of 400 BATSE SGRBs for which location errors were less than 10 degrees, with a sample of local galaxies, \citet{tanvir2005origin} were able to place a limit for the rate of short-duration gamma-ray bursts within $d\sim110$\,Mpc of $\sim$25\% of BATSE bursts.
Intriguingly, 
this analysis showed a positive cross-correlation at a $\sim$3$\sigma$ level.

The Gamma-ray Burst Monitor (GBM) instrument on the {Fermi Gamma-ray Space Telescope} ({Fermi}) has provided similar capabilities to BATSE for the past 10 years.
Again, although it has observed a large sample of bursts, positional accuracy is also much lower than {Swift}, at a few degrees \citep{connaughton2015}.

In this study, we revisit the cross-correlation analysis reported in \citet{tanvir2005origin} with an updated and more complete galaxy redshift catalogue (the 2MASS Redshift Survey (2MRS); \citep[]{Huchra2012})
and a larger sample of 782 bursts from combining both the {CGRO}/BATSE and {Fermi}/GBM catalogues.
We~require burst localisation error radii $<$10$^\circ$ and a $T_{90}\leq4$\,s,
and removed Fermi/GBM bursts that have been well-localised by other satellites.
The localisation errors also include estimates of systematic uncertainties following Model 2 for BATSE of \citet{briggs1999error} and \mbox{\citet{connaughton2015}} for GBM, each of which includes a core and tail systematic component.
The errors were assumed to follow a Fisher distribution \citep{briggs1999error}.
Distances to galaxies were taken from the HyperLEDA \citep{HyperLEDAIII} and~Cosmicflows~\citep{Cosmicflows3} compilations.

\begin{figure}[H]
	\centering
	\includegraphics[width=.9\textwidth]{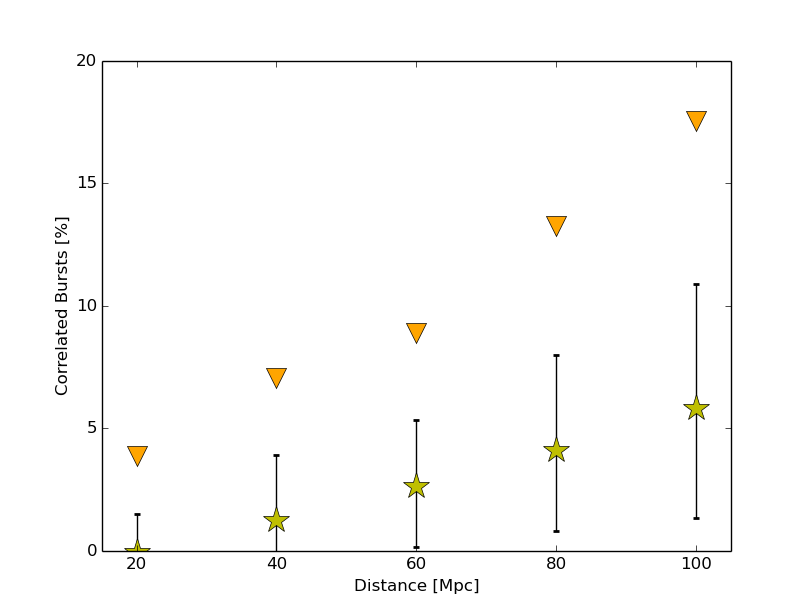}
	\caption{The correlated short-duration gamma-ray burst percentage (for $T_{90}<4$\,s) based on the $\Phi/\Phi_{0}$ spread determined for each $d_{gal}$ volume. The error bars correspond to 1$\sigma$ deviation from the observed correlation ratio, $\Phi_{obs}/\Phi_{0}$. The yellow stars represent the non-zero correlation percentage. Orange triangles are the 2$\sigma$ upper limits. 
  }
  \label{corfig}
\end{figure}

The correlation statistic, $\Phi$, matches each short-duration gamma-ray burst against every 2MRS galaxy within a given distance cut, summing up the likelihoods that a given burst would be found at the observed distance from the given galaxy if they were truly associated. Bursts were weighted inversely with their positional location error radii.
The galaxies were also weighted according to their absolute B-band luminosity, to provide some account for the likely higher rate of binary mergers in large galaxies and galaxies with ongoing or recent star formation (cf. \citep[]{Fong2013}).
The correlation statistic for the real bursts was then compared to a distribution of $\Phi$ values obtained for an artificial sample containing both randomly-distributed bursts, the average value of which we denote $\Phi_0$, as well as a fraction that were correlated with 2MRS galaxies.
This approach allows us to set limits on the fraction of correlated bursts in the real sample as a function of galaxy distance cut, shown in Figure~\ref{corfig}.

We conclude a 2$\sigma$ upper limit on the fraction of all BATSE and GBM short bursts within $\sim$100\,Mpc of $\lsim$17\%; this places an upper limit on the annual rate to be $\lsim$12 \textrm{y}$^{-1}$.

As can be seen from Figure~\ref{corfig}, there is no statistically-significant evidence for non-zero correlation at any distance. Although this conclusion differs from that of \citet{tanvir2005origin}, the two results are consistent within their $1\sigma$ error ranges, and the difference is primarily a consequence of updated samples of bursts and galaxies.

\subsection{A Search for Extragalactic SGR Giant Flares}

Distinguishing between transients related to flaring magnetars and those associated with merging NSs is difficult.
However, SGRs are relatively young and can be expected to trace the star formation, whereas merging compact binary systems may undergo a significant delay time or be kicked substantially away from their formation site \citep{bray2016}.
A nearby, $\lsim 10$-Mpc population of short gamma-ray transients that are associated with star-forming galaxies could indicate a population of extragalactic SGR GFs.

\begin{figure}[H]

\hbox{
\begin{subfigure}[h]{0.33\textwidth}
\includegraphics[height=7.5cm]{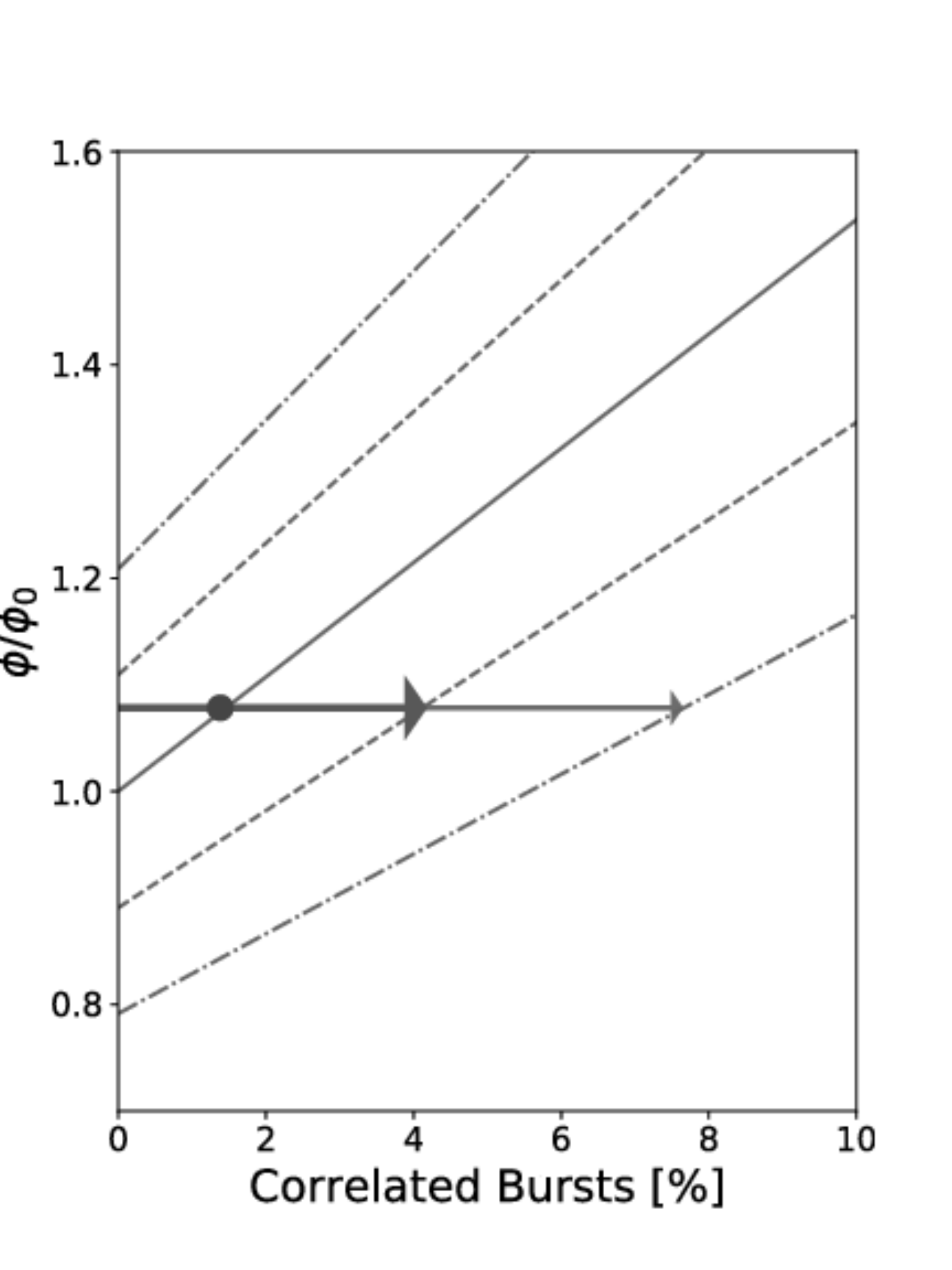}
\caption{}
\end{subfigure}
\begin{subfigure}[h]{0.66\textwidth}
\includegraphics[height=7.5cm]{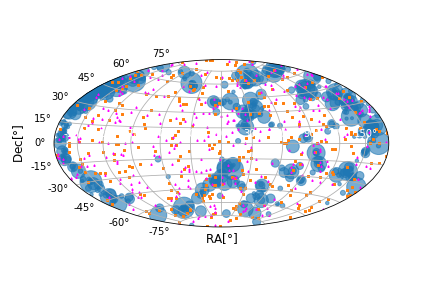}
\caption{}
\end{subfigure}

}
\vspace{-12pt}

\caption{(\textbf{a}) Ratio of the weighted correlation parameter $\Phi$ to $\Phi_0$, the value obtained for a random distribution of bursts,
as a function of the fraction of simulated bursts seeded by galaxy positions
(solid diagonal line). The galaxies were obtained from the compilation of \citet{karachentsev2013}, which is restricted to $d_{gal} = 11 \text{Mpc}$. The dashed lines in each case correspond to the $1\sigma$ and $2\sigma$ 
$\Phi/\Phi_{0}$ ranges determined by the spread in the simulation results.
The horizontal arrows are at the level of $\Phi/\Phi_0$ obtained for the real catalogue of BATSE and GBM $T_{90}<1$\,s bursts and indicate the plausible fraction that are correlated with nearby galaxies
(\textbf{b}) The observed galaxies (blue) within 11 Mpc with star formation rates from \citet{karachentsev2013}, shown alongside Burst and Transient Source Experiment (BATSE) (orange squares) and GBM
 (magenta triangles) SGRB detections. The area of each galactic point is scaled to reflect high star-forming environments.} 
\label{fig:k13_skymap}
\end{figure}

Within the BATSE and GBM sample of SGRBs a fraction of these bursts could be the result of SGR GFs.
Using Karachentsev {et al.'s} \citep{karachentsev2013} galaxy catalogue, which includes the star formation rate (SFR) for each galaxy, we computed the correlation for BATSE and GBM SGRBs with $T_{90}\leq 1$\,s with galaxies $<$11 Mpc weighted by their SFR.
The sky position for each of these bursts (orange squares for BATSE and magenta triangles for GBM) and the galaxies (blue disks) are shown in the right panel of Figure\,\ref{fig:k13_skymap}; the area of the galaxy marker is weighted by the SFR, so a larger area represents a higher SFR.
For~477 selected BATSE and Fermi GRBs in our sample, we find a 2$\sigma$ upper limit of $\sim8\%$, corresponding to $\lsim3\,{\rm y}^{-1}$, for the fraction of bursts that could be correlated with these nearby, high SFR galaxies (see~the left panel of Figure \ref{fig:k13_skymap}).
Once again, we note that the absence of $d<11$-Mpc galaxies in the error regions of any {Swift}-detected bursts, confirming that this limit is a hard one.

A number of previous studies constrained the fraction of SGR GFs in SGRB catalogues.
A~sample of 76 well-detected BATSE SGRBs was analysed by \citet{lazzati2005soft}, who concluded that only three were consistent with having black body spectra, and so candidates for extra-galactic GFs, albeit the durations of these candidates were $>1$\,s and longer than the GFs observed in the Milky Way~system.

\citet{popov2006} argued that the rate of GFs observed in the Milky Way system would lead to the expectation that 15--25 SGR GFs from four galaxies (M82, NGC\,253, NGC\,4945, and M83) should have been detected during the life of the BATSE instrument.
They noted that some BATSE SGRBs had positions consistent with these galaxies, but when looked at in detail, concluded that their spectra and light curves were not as expected for GFs.
Furthermore, over the current elapsed mission time, Swift has made no such detections arising from M82 or similar candidates such as NGC 253, NGC 4945, and~M83, despite having a somewhat reduced trigger threshold for many of these nearby galaxies.

An upper limit on the fraction of SGR GFs in the ($T_{90}<2$\,s) SGRB population at $<$15\% was made by \citet{nakar2006} using a sample of six well-localised SGRBs.
Similarly, using 47 Inter-Planetary Network (IPN) localised SGRBs, \citet{ofek2007} checked for coincidence between bright and star-forming galaxies within 20 Mpc and the SGRB IPN sky position annulus for each burst.
A single match between GRB\,000420B and M74 was found, although this is likely to be a chance coincidence.
By assuming an upper limit cut-off for the SGR GF isotropic energy distribution of $<$10$^{47}$\,erg, an upper limit of $<$16\% was found for the fraction of SGR GFs in the SGRB population.
\citet{ofek2007} placed a lower limit on the fraction at $1\%$ based on the Galactic SGR GF rate.
This range for the fraction of 1--15\%, based on statistical analysis, was reported by \citet{hurley2011}.

The upper limit for the fraction of SGR GFs in the SGRB population was estimated to be $<$7\% by \citet{tikhomirova2010} from analysis of SGRBs with relatively small localisation regions based on BATSE and IPN detections. More recently, \citet{svinkin2015search} found a similar limit of $<$8\%;
they considered the evidence for extragalactic SGR GFs amongst 16 years of well-localised Konus-{WIND
} SGRBs and found no compelling cases apart from the GRBs\,051103 and 070201, discussed in Section~\ref{sec:ipn}.

\vspace{-12pt}

\section{IPN-Observed SGRBs}
\label{sec:ipn}

The Inter-Planetary Network (IPN) is the name given to a collaborative group of GRB detectors on various satellites and probes. Some of these are dedicated GRB missions, but others piggy-back on other missions.
It has been operational through most of the past four decades and has nodes both at the Earth and elsewhere in the Solar System.
The location of a burst can be accurately constrained by measuring the arrival time across several IPN satellites; the precision of which
depends on the baseline separation 
\citep{cline1980,evans1980location,svinkin2015search}.
The long operational life of the network, and all-sky coverage, has made it particularly successful at finding rare and bright GRB events \citep{cline1981precise}.
In fact, the IPN has detected all three SGR giant flares originating in the MW/LMC \citep{svinkin2015search}.

IPN detections have helped provide low-z host candidates for several bursts, notably M31 for GRB\,070201 \citep{Perley2007gcn}, M81 for GRB\,051103 \citep{hurley2010new}, and NGC\,3313 for GRB\,150906B \citep{Levan2015gcn}.
The former two were particularly bright events, adding to the case of their being at short distances.
If these host associations are correct, then the isotropic equivalent energy for GRB\,051103 and GRB\,070201 would be $7.5\times10^{46}$\,erg and $1.5\times10^{45}$\,erg, respectively \citep{hurley2010new}.
These are interestingly close to the fluence of GRB\,170817A;
however, the lack of coincident gravitational wave detections makes it unlikely that these bursts originated from compact binary mergers \citep{abbott2008implications,abadie2012,abbott2017search}.
Based on the proximity of the potential host galaxies, it is more likely that these detections correspond to giant flares from soft gamma-ray repeaters (SGRs), particularly for GRBs 051103 and 070201.

\section{Discussion and Conclusions}

We have estimated the number of Swift gamma-ray bursts with $T_{90}<4$\,s that could be associated with galaxies within the Advanced-LIGO horizon.
A duration cut at 4\,s was adopted considering that events similar
to GRB\,170817A may be typically rather longer than traditional
short-duration GRBs.
We searched for any with minimum separation distance (or impact parameter $b$) of the burst from a galaxy of $b<200$\,kpc
to allow for the possibility of NSNS/NSBH binaries being kicked to large distances from their hosts in some cases.
In roughly two all-sky years of {Swift} observations, we find eight host candidates within $200$\,Mpc.
The majority of these are likely to be chance associations, which
suggests a maximum detectable SGRB all-sky rate of $<4 \text{ y}^{-1}$ at this distance. Only if typical kick distances are very large ($\gg200$\,kpc) would it become difficult to identify the potential hosts for these SGRBs due to the large separation, which would potentially allow a higher rate.

Based on the impact parameter analysis of {Swift} bursts, the corresponding volumetric rate density is $R<$\,$120\textrm{ Gpc}^{-3} \,\textrm{y}^{-1}$. This falls below the $1\sigma$ range estimated for NSNS mergers by \mbox{\citet{abbott2017gw170817}} of $R = 1540^{+3200}_{-1220} \textrm{ Gpc}^{-3} \,\textrm{y}^{-1} $ based on the the single detection of GW170817 during the O1 and O2 LIGO science runs. Thus, if the true merger rate is as high as this, then only a small fraction of future GW detections is likely to be accompanied by detectable gamma-ray flashes. That could be understood as a consequence of anisotropic gamma-ray emission, since our line of sight was within $30^{\circ}$ of the primary jet axis in the case of GW170817.
On the other hand, estimates of the NSNS merger rate based on the small sample of known Milky Way double neutron stars continue to point to lower figures, albeit also with large uncertainties (e.g., $R\sim50\,{\rm Gpc}^{-3}\,{\rm y}^{-1}$ in \citep[]{Chruslinska2018}).

The strongest candidate for a low-z short GRB remains GRB 050906 \cite{Levan08GRB050906}, whose BAT error box contains the galaxy IC 0328 at
132 Mpc. At this distance, its isotropic equivalent energy would be $E_{iso} = 1.9 \times 10^{46}$ erg (15-150 keV) (the bolometric correction likely yields a total energy a factor of a few higher than this), which lies in the same regime as GW170817/GRB170817A. The non-detection of a clear X-ray afterglow at the time of GRB 050906 was regarded as a likely indication that it was not
a local event. However, the experience with GW170817/GRB170817A shows that off-axis afterglow emission can rise at later times at all wavelengths \cite{troja2017,hallinan2017,lyman2018} at levels undetectable to the Swift-XRT \cite{evans2017}. Hence, GRB 050906 could well be associated with a binary merger within IC 0328, which was viewed away from its axis, although deep optical and near-IR imaging places strong limits on any emission from any accompanying kilonova.

For bursts with a poorer localisation, from {CGRO}/BATSE and {Fermi}/GBM, we follow the statistical method of \citet{tanvir2005origin} to measure the correlation of burst positions with those of nearby galaxies.
We have found upper limits on the fraction of gamma-ray bursts with $T_{90}<4$\,s detected by BATSE and GBM that could be associated with galaxies within $100$\,Mpc.
This analysis results in a weaker constraint, setting a $2\sigma$ upper limit for the fraction of bursts at $\lsim17\%$ (see Figure \ref{corfig}).

An extragalactic population of SGR GFs, similar to those observed in the Milky Way system, within 10--25\,Mpc would appear as low-luminosity SGRBs.
We have tailored a study specifically to constrain the fraction of SGR GF events masquerading as SGRBs, 
using galaxies within 11 Mpc, weighting them by their star formation rate \citep{karachentsev2013}, correlating them with bursts with a $T_{90}<1$ s (as~shown in Figure \ref{fig:k13_skymap}).
This analysis places a $2\sigma$ upper limit on the fraction of such SGRBs that are due to SGR GFs within $11$\,Mpc of $<$3\,{\rm y}$^{-1}$.
This limit is supported by the lack of {Swift} identified SGR GF events in nearby galaxies.
In turn, given the small number statistics, it is reasonably consistent with the observation of three GFs in the Milky Way system in $\sim$40\,y, taking into account that the overall star formation rate in this volume is about 50$\times$ that of the Milky Way system (e.g., \citep[]{licquia2015}).

Another potential source of low-luminosity GRB-like events results from resonant shattering flares of neutron stars shortly before their merger in a binary system.
These gamma-rays are emitted isotropically from the source and will usually not be accompanied by an SGRB beamed towards the observer.
In rarer cases where the SGRB is favourably aligned, these bursts of gamma-rays could appear as a precursor to the SGRB.
These short duration bursts of gamma-rays would appear similar to a population of SGR GFs, but would follow the same host distribution and offsets as the SGRB population.
As it is unlikely that an SGR would have any offset from its host galaxy, due to their short lifetime and association with star-forming regions, the candidates with $T_{90}\lsim0.5$\,s listed in Table \ref{tab:impact} could include a population of RSFs or other precursor type transients.

Finally, we reiterate our main conclusion for the prospects of coincident gamma-ray signals with GW merger events found in the upcoming runs of the advanced generation gravitational wave detectors.
The predicted NSNS detection rate during these runs remains very uncertain, but estimates range up to 50--80 per year \citep{abbottLRR}. If the true rate of mergers is as high as this, then our results suggest that only a small percentage, $<$10\%, is likely to exhibit prompt gamma-ray flashes.

%%%%%%%%%%%%%%%%%%%%%%%%%%%%%%%%%%%%%%%%%%
\vspace{6pt}

%%%%%%%%%%%%%%%%%%%%%%%%%%%%%%%%%%%%%%%%%%
\authorcontributions{ S.F.M., N.R.T., and G.P.L wrote the paper. N.R.T. conceived of the original analysis. S.F.M., N.R.T., and G.P.L. contributed equally to the revised analysis noted in this article. A.J.L. and D.T. contributed comments and analysis to assist in the writing of this manuscript. Breakdown: Conceptualization, S.M., N.T., and G.L.; formal analysis, S.M., N.T., and G.L.; investigation, S.M., N.T., and G.L.; methodology, S.M., N.T., and G.L.; supervision, N.T.; validation, N.T., G.L., A.L., and D.T.; writing, original draft, S.M., N.T., G.L., A.L., and D.T.}

\acknowledgments{The authors thank Andrew Blain for useful discussions. We would like to extend our gratitude to the reviewers of this paper for their useful feedback and comments. We acknowledge the usage of the following databases: HyperLEDA; Extragalactic Distance Database; NASA/IPAC Extragalactic Database; Two Mass Redshift Survey. For the images used in Figure \ref{fig:SGRBs}, we acknowledge the usage of the Digitized Sky Survey produced at the Space Telescope Science Institute under U.S. Government Grant NAG W-2166. The images of these surveys are based on photographic data obtained using the Oschin Schmidt Telescope on Palomar Mountain and the U.K. Schmidt Telescope. The plates were processed into the present compressed digital form with the permission of these institutions. 
S.F.M. is supported by a PhD studentship funded by
the College of Science and Engineering at the University of Leicester; G.P.L. is supported by STFC grants; N.R.T. and A.J.L. acknowledge support through ERC Grant 725246 TEDE.
}
%%%%%%%%%%%%%%%%%%%%%%%%%%%%%%%%%%%%%%%%%%
%%%%%%%%%%%%%%%%%%%%%%%%%%%%%%%%%%%%%%%%%%
\conflictofinterests{The authors declare no conflict of interest.}

%%%%%%%%%%%%%%%%%%%%%%%%%%%%%%%%%%%%%%%%%%
%% optional
\abbreviations{The following abbreviations are used in this manuscript:\\
\noindent
\begin{tabular}{@{}ll}
SGRB&Short-Duration Gamma-ray Burst\\
NSNS& Binary Neutron Stars\\
NS&Neutron Star\\
BHNS&Black Hole-Neutron Star Pair\\

\end{tabular}

\noindent
\begin{tabular}{@{}ll}
SGR& Soft Gamma-ray Tepeater\\
GF& Giant Flare\\
RSF& Resonant Shattering Flare\\
SFR& Star Formation Rate\\
MW&Milky Way\\
LMC& Large Magellanic Cloud\\
DSS& Digitized Sky Survey\\
2MASS& Two Micron All-Sky Survey\\
VLT& Very Large Telescope\\
VIMOS& Visible Multi Object Spectrograph\\
GBM& [Fermi] Gamma-ray Burst Monitor\\
XRT& [Swift] X-Ray Telescope\\
IPN& Inter-Planetary Network
\end{tabular}}

\bibliographystyle{mdpi}
\bibliography{ms}

\begin{thebibliography}{-------}
\providecommand{\natexlab}[1]{#1}

\bibitem[{Kouveliotou} \em{et~al.}(1993){Kouveliotou}, {Meegan}, {Fishman}, and
  e.a.]{kouveliotou1993}
{Kouveliotou}, C.; {Meegan}, C.A.; {Fishman}, G.J.; e.a..
\newblock {Identification of Two Classes of Gamma-Ray Bursts}.
\newblock {\em \apj} {\bf 1993}, {\em 413},~L101.

\bibitem[{Bromberg} \em{et~al.}(2013){Bromberg}, {Nakar}, {Piran}, and
  {Sari}]{Bromberg2013}
{Bromberg}, O.; {Nakar}, E.; {Piran}, T.; {Sari}, R.
\newblock {Short versus Long and Collapsars versus Non-collapsars: A
  Quantitative Classification of Gamma-Ray Bursts}.
\newblock {\em \apj} {\bf 2013}, {\em 764},~179.

\bibitem[{Eichler} \em{et~al.}(1989){Eichler}, {Livio}, {Piran}, and
  {Schramm}]{eichler1989}
{Eichler}, D.; {Livio}, M.; {Piran}, T.; {Schramm}, D.N.
\newblock {Nucleosynthesis, neutrino bursts and {\ensuremath{\gamma}}-rays from
  coalescing neutron stars}.
\newblock {\em \nat} {\bf 1989}, {\em 340},~126--128.

\bibitem[{Narayan} \em{et~al.}(1992){Narayan}, {Paczynski}, and
  {Piran}]{narayan1992}
{Narayan}, R.; {Paczynski}, B.; {Piran}, T.
\newblock {Gamma-Ray Bursts as the Death Throes of Massive Binary Stars}.
\newblock {\em \apj} {\bf 1992}, {\em 395},~L83.

\bibitem[{Mochkovitch} \em{et~al.}(1993){Mochkovitch}, {Hernanz}, {Isern}, and
  {Martin}]{mochkovitch1993}
{Mochkovitch}, R.; {Hernanz}, M.; {Isern}, J.; {Martin}, X.
\newblock {Gamma-ray bursts as collimated jets from neutron star/black hole
  mergers}.
\newblock {\em \nat} {\bf 1993}, {\em 361},~236--238.

\bibitem[{Bogomazov} \em{et~al.}(2007){Bogomazov}, {Lipunov}, and
  {Tutukov}]{bogomazov2007}
{Bogomazov}, A.I.; {Lipunov}, V.M.; {Tutukov}, A.V.
\newblock {Evolution of close binaries and gamma-ray bursts}.
\newblock {\em Astronomy Reports} {\bf 2007}, {\em 51},~308--317.

\bibitem[Roberts \em{et~al.}(2011)Roberts, Kasen, Lee, and
  Ramirez-Ruiz]{roberts2011electromagnetic}
Roberts, L.F.; Kasen, D.; Lee, W.H.; Ramirez-Ruiz, E.
\newblock Electromagnetic transients powered by nuclear decay in the tidal
  tails of coalescing compact binaries.
\newblock {\em \apjl} {\bf 2011}, {\em 736},~L21.

\bibitem[Giacomazzo \em{et~al.}(2012)Giacomazzo, Perna, Rezzolla, Troja, and
  Lazzati]{giacomazzo2012compact}
Giacomazzo, B.; Perna, R.; Rezzolla, L.; Troja, E.; Lazzati, D.
\newblock Compact binary progenitors of short gamma-ray bursts.
\newblock {\em \apjl} {\bf 2012}, {\em 762},~L18.

\bibitem[Paschalidis(2017)]{paschalidis2017general}
Paschalidis, V.
\newblock General relativistic simulations of compact binary mergers as engines
  for short gamma-ray bursts.
\newblock {\em Classical and Quantum Gravity} {\bf 2017}, {\em 34},~084002.

\bibitem[Berger(2014)]{berger2014short}
Berger, E.
\newblock Short-duration gamma-ray bursts.
\newblock {\em \araa} {\bf 2014}, {\em 52},~43--105.

\bibitem[Fong and Berger(2013)]{fong2013locations}
Fong, W.F.; Berger, E.
\newblock The locations of short gamma-ray bursts as evidence for compact
  object binary progenitors.
\newblock {\em \apj} {\bf 2013}, {\em 776},~18.

\bibitem[Tunnicliffe \em{et~al.}(2013)Tunnicliffe, Levan, Tanvir, and
  e.a.]{tunnicliffe2013nature}
Tunnicliffe, R.L.; Levan, A.J.; Tanvir, N.R.; e.a..
\newblock On the nature of the ‘hostless’ short GRBs.
\newblock {\em \mnras} {\bf 2013}, {\em 437},~1495--1510.

\bibitem[{Bray} and {Eldridge}(2016)]{bray2016}
{Bray}, J.C.; {Eldridge}, J.J.
\newblock {Neutron star kicks and their relationship to supernovae ejecta
  mass}.
\newblock {\em \mnras} {\bf 2016}, {\em 461},~3747--3759.

\bibitem[{Tanvir} \em{et~al.}(2013){Tanvir}, {Levan}, and
  {Fruchter}]{tanvir2013}
{Tanvir}, N.R.; {Levan}, A.J.; {Fruchter}, A.S.e.
\newblock {A `kilonova' associated with the short-duration
  {\ensuremath{\gamma}}-ray burst GRB 130603B}.
\newblock {\em \nat} {\bf 2013}, {\em 500},~547--549.

\bibitem[Berger \em{et~al.}(2013)Berger, Fong, and Chornock]{berger2013r}
Berger, E.; Fong, W.; Chornock, R.
\newblock An r-process kilonova associated with the short-hard GRB 130603B.
\newblock {\em \apjl} {\bf 2013}, {\em 774},~L23.

\bibitem[Jin \em{et~al.}(2015)Jin, Li, Cano, and e.a.]{jin2015light}
Jin, Z.P.; Li, X.; Cano, Z.; e.a..
\newblock The Light Curve of the Macronova Associated With the Long--short
  Burst GRB 060614.
\newblock {\em \apjl} {\bf 2015}, {\em 811},~L22.

\bibitem[{Gompertz} \em{et~al.}(2018){Gompertz}, {Levan}, {Tanvir}, {Hjorth},
  {Covino}, {Evans}, {Fruchter}, {Gonz{\'a}lez-Fern{\'a}ndez}, {Jin}, {Lyman},
  and e.a.]{gompertz2018}
{Gompertz}, B.P.; {Levan}, A.J.; {Tanvir}, N.R.; {Hjorth}, J.; {Covino}, S.;
  {Evans}, P.A.; {Fruchter}, A.S.; {Gonz{\'a}lez-Fern{\'a}ndez}, C.; {Jin},
  Z.P.; {Lyman}, J.D.; e.a..
\newblock {The Diversity of Kilonova Emission in Short Gamma-Ray Bursts}.
\newblock {\em \apj} {\bf 2018}, {\em 860},~62.

\bibitem[{Abbott} \em{et~al.}(2017){Abbott}, {Abbott}, {Abbott}, {Acernese},
  {Ackley}, {Adams}, {Adams}, {Addesso}, {Adhikari}, {Adya}, and
  e.a.]{abbott2017}
{Abbott}, B.P.; {Abbott}, R.; {Abbott}, T.D.; {Acernese}, F.; {Ackley}, K.;
  {Adams}, C.; {Adams}, T.; {Addesso}, P.; {Adhikari}, R.X.; {Adya}, V.B.;
  e.a..
\newblock {Gravitational Waves and Gamma-Rays from a Binary Neutron Star
  Merger: GW170817 and GRB 170817A}.
\newblock {\em \apj} {\bf 2017}, {\em 848},~L13.

\bibitem[{Beniamini} \em{et~al.}(2018){Beniamini}, {Petropoulou}, {Barniol
  Duran}, and {Giannios}]{beniamini2018b}
{Beniamini}, P.; {Petropoulou}, M.; {Barniol Duran}, R.; {Giannios}, D.
\newblock {A lesson from GW170817: most neutron star mergers result in tightly
  collimated successful GRB jets}.
\newblock {\em \mnras} {\bf 2018}, p. 2945.

\bibitem[Tanvir \em{et~al.}(2005)Tanvir, Chapman, Levan, and
  Priddey]{tanvir2005origin}
Tanvir, N.R.; Chapman, R.; Levan, A.J.; Priddey, R.S.
\newblock An origin in the local Universe for some short $\gamma$-ray bursts.
\newblock {\em \nat} {\bf 2005}, {\em 438},~991.

\bibitem[Chapman \em{et~al.}(2009)Chapman, Priddey, and
  Tanvir]{chapman2009short}
Chapman, R.; Priddey, R.S.; Tanvir, N.R.
\newblock Short gamma-ray bursts from SGR giant flares and neutron star
  mergers: two populations are better than one.
\newblock {\em \mnras} {\bf 2009}, {\em 395},~1515--1522.

\bibitem[{Rowlinson} \em{et~al.}(2010){Rowlinson}, {Wiersema}, {Levan},
  {Tanvir}, {O'Brien}, {Rol}, {Hjorth}, {Th{\"o}ne}, {de Ugarte Postigo},
  {Fynbo}, and e.a.]{rowlinson2010unusual}
{Rowlinson}, A.; {Wiersema}, K.; {Levan}, A.J.; {Tanvir}, N.R.; {O'Brien},
  P.T.; {Rol}, E.; {Hjorth}, J.; {Th{\"o}ne}, C.C.; {de Ugarte Postigo}, A.;
  {Fynbo}, J.P.U.; e.a..
\newblock {Discovery of the afterglow and host galaxy of the low-redshift short
  GRB 080905A}.
\newblock {\em \mnras} {\bf 2010}, {\em 408},~383--391.

\bibitem[{de Ugarte Postigo} \em{et~al.}(2006){de Ugarte Postigo},
  {Castro-Tirado}, {Guziy}, {Gorosabel}, {J{\'o}hannesson}, {Aloy}, {McBreen},
  {Lamb}, {Benitez}, {Jel{\'\i}nek}, and e.a.]{Gorosabel2006}
{de Ugarte Postigo}, A.; {Castro-Tirado}, A.J.; {Guziy}, S.; {Gorosabel}, J.;
  {J{\'o}hannesson}, G.; {Aloy}, M.A.; {McBreen}, S.; {Lamb}, D.Q.; {Benitez},
  N.; {Jel{\'\i}nek}, M.; e.a..
\newblock {GRB 060121: Implications of a Short-/Intermediate-Duration
  {$\gamma$}-Ray Burst at High Redshift}.
\newblock {\em \apjl} {\bf 2006}, {\em 648},~L83--L87.

\bibitem[{Selsing} \em{et~al.}(2018){Selsing}, {Kr{\"u}hler}, {Malesani},
  {D'Avanzo}, {Schulze}, {Vergani}, {Palmerio}, {Japelj}, {Milvang-Jensen},
  {Watson}, and e.a.]{selsing2018host}
{Selsing}, J.; {Kr{\"u}hler}, T.; {Malesani}, D.; {D'Avanzo}, P.; {Schulze},
  S.; {Vergani}, S.D.; {Palmerio}, J.; {Japelj}, J.; {Milvang-Jensen}, B.;
  {Watson}, D.; e.a..
\newblock The host galaxy of the short GRB 111117A at z= 2.211-Impact on the
  short GRB redshift distribution and progenitor channels.
\newblock {\em \aap} {\bf 2018}, {\em 616},~A48.

\bibitem[{Lamb} and {Kobayashi}(2017)]{lamb2017}
{Lamb}, G.P.; {Kobayashi}, S.
\newblock {Electromagnetic counterparts to structured jets from gravitational
  wave detected mergers}.
\newblock {\em \mnras} {\bf 2017}, {\em 472},~4953--4964.

\bibitem[{Jin} \em{et~al.}(2018){Jin}, {Li}, {Wang}, {Wang}, {He}, {Yuan},
  {Zhang}, {Zou}, {Fan}, and {Wei}]{jin2018}
{Jin}, Z.P.; {Li}, X.; {Wang}, H.; {Wang}, Y.Z.; {He}, H.N.; {Yuan}, Q.;
  {Zhang}, F.W.; {Zou}, Y.C.; {Fan}, Y.Z.; {Wei}, D.M.
\newblock {Short GRBs: Opening Angles, Local Neutron Star Merger Rate, and
  Off-axis Events for GRB/GW Association}.
\newblock {\em \apj} {\bf 2018}, {\em 857},~128.

\bibitem[{Kathirgamaraju} \em{et~al.}(2018){Kathirgamaraju}, {Barniol Duran},
  and {Giannios}]{kathirgamaraju2018}
{Kathirgamaraju}, A.; {Barniol Duran}, R.; {Giannios}, D.
\newblock {Off-axis short GRBs from structured jets as counterparts to GW
  events}.
\newblock {\em \mnras} {\bf 2018}, {\em 473},~L121--L125.

\bibitem[{Mooley} \em{et~al.}(2018){Mooley}, {Deller}, {Gottlieb}, {Nakar},
  {Hallinan}, {Bourke}, {Frail}, {Horesh}, {Corsi}, and
  {Hotokezaka}]{mooley2018}
{Mooley}, K.P.; {Deller}, A.T.; {Gottlieb}, O.; {Nakar}, E.; {Hallinan}, G.;
  {Bourke}, S.; {Frail}, D.A.; {Horesh}, A.; {Corsi}, A.; {Hotokezaka}, K.
\newblock {Superluminal motion of a relativistic jet in the neutron-star merger
  GW170817}.
\newblock {\em \nat} {\bf 2018}, {\em 561},~355--359.

\bibitem[{van Eerten} \em{et~al.}(2018){van Eerten}, {Ryan}, {Ricci},
  {Burgess}, {Wieringa}, {Piro}, {Cenko}, and {Sakamoto}]{troja2018}
{van Eerten}, E.T.H.; {Ryan}, G.; {Ricci}, R.; {Burgess}, J.M.; {Wieringa}, M.;
  {Piro}, L.; {Cenko}, S.B.; {Sakamoto}, T.
\newblock {A year in the life of GW170817: the rise and fall of a structured
  jet from a binary neutron star merger}.
\newblock {\em ArXiv e-prints} {\bf 2018}, p. arXiv:1808.06617.

\bibitem[{Lazzati} \em{et~al.}(2017){Lazzati}, {L{\'o}pez-C{\'a}mara},
  {Cantiello}, {Morsony}, {Perna}, and {Workman}]{lazzati2017}
{Lazzati}, D.; {L{\'o}pez-C{\'a}mara}, D.; {Cantiello}, M.; {Morsony}, B.J.;
  {Perna}, R.; {Workman}, J.C.
\newblock {Off-axis Prompt X-Ray Transients from the Cocoon of Short Gamma-Ray
  Bursts}.
\newblock {\em \apj} {\bf 2017}, {\em 848},~L6.

\bibitem[{Kasliwal} \em{et~al.}(2017){Kasliwal}, {Nakar}, {Singer}, {Kaplan},
  {Cook}, {Van Sistine}, {Lau}, {Fremling}, {Gottlieb}, {Jencson}, and
  e.a.]{kasliwal2017}
{Kasliwal}, M.M.; {Nakar}, E.; {Singer}, L.P.; {Kaplan}, D.L.; {Cook}, D.O.;
  {Van Sistine}, A.; {Lau}, R.M.; {Fremling}, C.; {Gottlieb}, O.; {Jencson},
  J.E.; e.a..
\newblock {Illuminating gravitational waves: A concordant picture of photons
  from a neutron star merger}.
\newblock {\em Science} {\bf 2017}, {\em 358},~1559--1565.

\bibitem[{Gottlieb} \em{et~al.}(2018){Gottlieb}, {Nakar}, and
  {Hotokezaka}]{gottlieb2018}
{Gottlieb}, O.; {Nakar}, E.and~{Piran}, T.; {Hotokezaka}, K.
\newblock {A cocoon shock breakout as the origin of the
  {\ensuremath{\gamma}}-ray emission in GW170817}.
\newblock {\em \mnras} {\bf 2018}, {\em 479},~588--600.

\bibitem[{Lamb} and {Kobayashi}(2018)]{lamb2018}
{Lamb}, G.P.; {Kobayashi}, S.
\newblock {GRB 170817A as a jet counterpart to gravitational wave trigger GW
  170817}.
\newblock {\em \mnras} {\bf 2018}, {\em 478},~733--740.

\bibitem[{Troja} \em{et~al.}(2010){Troja}, {Rosswog}, and {Gehrels}]{troja2010}
{Troja}, E.; {Rosswog}, S.; {Gehrels}, N.
\newblock {Precursors of Short Gamma-ray Bursts}.
\newblock {\em \apj} {\bf 2010}, {\em 723},~1711--1717.

\bibitem[{Kochanek}(1992)]{kochanek1992}
{Kochanek}, C.S.
\newblock {Coalescing Binary Neutron Stars}.
\newblock {\em \apj} {\bf 1992}, {\em 398},~234.

\bibitem[{Tsang}(2013)]{tsang2013}
{Tsang}, D.
\newblock {Shattering Flares during Close Encounters of Neutron Stars}.
\newblock {\em \apj} {\bf 2013}, {\em 777},~103.

\bibitem[{Kyutoku} \em{et~al.}(2014){Kyutoku}, {Ioka}, and
  {Shibata}]{kyutoku2014}
{Kyutoku}, K.; {Ioka}, K.; {Shibata}, M.
\newblock {Ultrarelativistic electromagnetic counterpart to binary neutron star
  mergers}.
\newblock {\em \mnras} {\bf 2014}, {\em 437},~L6--L10.

\bibitem[{Metzger} and {Zivancev}(2016)]{metzger2016}
{Metzger}, B.D.; {Zivancev}, C.
\newblock {Pair fireball precursors of neutron star mergers}.
\newblock {\em \mnras} {\bf 2016}, {\em 461},~4435--4440.

\bibitem[{Kouveliotou} \em{et~al.}(1999){Kouveliotou}, {Strohmayer}, {Hurley},
  {van Paradijs}, {Finger}, {Dieters}, {Woods}, {Thompson}, and
  {Duncan}]{kouveliotou1999}
{Kouveliotou}, C.; {Strohmayer}, T.; {Hurley}, K.; {van Paradijs}, J.;
  {Finger}, M.H.; {Dieters}, S.; {Woods}, P.; {Thompson}, C.; {Duncan}, R.C.
\newblock {Discovery of a Magnetar Associated with the Soft Gamma Repeater SGR
  1900+14}.
\newblock {\em \apjl} {\bf 1999}, {\em 510},~L115--L118.

\bibitem[{Palmer} \em{et~al.}(2005){Palmer}, {Barthelmy}, {Gehrels}, {Kippen},
  {Cayton}, {Kouveliotou}, {Eichler}, {Wijers}, {Woods}, {Granot}, and
  e.a.]{palmer2005}
{Palmer}, D.M.; {Barthelmy}, S.; {Gehrels}, N.; {Kippen}, R.M.; {Cayton}, T.;
  {Kouveliotou}, C.; {Eichler}, D.; {Wijers}, R.A.M.J.; {Woods}, P.M.;
  {Granot}, J.; e.a..
\newblock {A giant {\ensuremath{\gamma}}-ray flare from the magnetar SGR 1806 -
  20}.
\newblock {\em \nat} {\bf 2005}, {\em 434},~1107--1109.

\bibitem[{Hurley} \em{et~al.}(2005){Hurley}, {Boggs}, {Smith}, {Duncan}, {Lin},
  {Zoglauer}, {Krucker}, {Hurford}, {Hudson}, {Wigger}, and e.a.]{hurley2005}
{Hurley}, K.; {Boggs}, S.E.; {Smith}, D.M.; {Duncan}, R.C.; {Lin}, R.;
  {Zoglauer}, A.; {Krucker}, S.; {Hurford}, G.; {Hudson}, H.; {Wigger}, C.;
  e.a..
\newblock {An exceptionally bright flare from SGR 1806-20 and the origins of
  short- duration {\ensuremath{\gamma}}-ray bursts}.
\newblock {\em \nat} {\bf 2005}, {\em 434},~1098--1103.

\bibitem[{Hurley}(2011)]{hurley2011}
{Hurley}, K.
\newblock {The short gamma-ray burst - SGR giant flare connection}.
\newblock {\em Advances in Space Research} {\bf 2011}, {\em 47},~1337--1340.

\bibitem[{Bibby} \em{et~al.}(2008){Bibby}, {Crowther}, {Furness}, and
  {Clark}]{bibby2008}
{Bibby}, J.L.; {Crowther}, P.A.; {Furness}, J.P.; {Clark}, J.S.
\newblock {A downward revision to the distance of the 1806-20 cluster and
  associated magnetar from Gemini Near-Infrared Spectroscopy}.
\newblock {\em \mnras} {\bf 2008}, {\em 386},~L23--L27.

\bibitem[{Cline} \em{et~al.}(1980){Cline}, {Desai}, {Pizzichini}, {Teegarden},
  {Evans}, {Klebesadel}, {Laros}, {Hurley}, {Niel}, and {Vedrenne}]{cline1980}
{Cline}, T.L.; {Desai}, U.D.; {Pizzichini}, G.; {Teegarden}, B.J.; {Evans},
  W.D.; {Klebesadel}, R.W.; {Laros}, J.G.; {Hurley}, K.; {Niel}, M.;
  {Vedrenne}, G.
\newblock {Detection of a fast, intense and unusual gamma-ray transient.}
\newblock {\em \apj} {\bf 1980}, {\em 237},~L1--L5.

\bibitem[{Thompson} and {Duncan}(2001)]{thompson2001}
{Thompson}, C.; {Duncan}, R.C.
\newblock {The Giant Flare of 1998 August 27 from SGR 1900+14. II. Radiative
  Mechanism and Physical Constraints on the Source}.
\newblock {\em \apj} {\bf 2001}, {\em 561},~980--1005.

\bibitem[{Gehrels} \em{et~al.}(2004){Gehrels}, {Chincarini}, {Giommi}, {Mason},
  {Nousek}, {Wells}, {White}, {Barthelmy}, {Burrows}, {Cominsky}, and
  e.a.]{gehrels2004swift}
{Gehrels}, N.; {Chincarini}, G.; {Giommi}, P.; {Mason}, K.O.; {Nousek}, J.A.;
  {Wells}, A.A.; {White}, N.E.; {Barthelmy}, S.D.; {Burrows}, D.N.; {Cominsky},
  L.R.; e.a..
\newblock The Swift gamma-ray burst mission.
\newblock {\em \apj} {\bf 2004}, {\em 611},~1005.

\bibitem[{Barthelmy} \em{et~al.}(2005){Barthelmy}, {Barbier}, {Cummings},
  {Fenimore}, {Gehrels}, {Hullinger}, {Krimm}, {Markwardt}, {Palmer},
  {Parsons}, and e.a.]{barthelmy2005burst}
{Barthelmy}, S.D.; {Barbier}, L.M.; {Cummings}, J.R.; {Fenimore}, E.E.;
  {Gehrels}, N.; {Hullinger}, D.; {Krimm}, H.A.; {Markwardt}, C.B.; {Palmer},
  D.M.; {Parsons}, A.; e.a..
\newblock The Burst Alert Telescope (BAT) on the SWIFT Midex mission.
\newblock {\em Space Science Reviews} {\bf 2005}, {\em 120},~143--164.

\bibitem[{Zhang} \em{et~al.}(2006){Zhang}, {Fan}, {Dyks}, {Kobayashi},
  {M{\'e}sz{\'a}ros}, {Burrows}, {Nousek}, and {Gehrels}]{zhang2006physical}
{Zhang}, B.; {Fan}, Y.Z.; {Dyks}, J.; {Kobayashi}, S.; {M{\'e}sz{\'a}ros}, P.;
  {Burrows}, D.N.; {Nousek}, J.A.; {Gehrels}, N.
\newblock Physical Processes Shaping Gamma-Ray Burst X-Ray Afterglow Light
  Curves: Theoretical Implications from the Swift X-Ray Telescope Observations.
\newblock {\em \apj} {\bf 2006}, {\em 642},~354.

\bibitem[{Brown} \em{et~al.}(2009){Brown}, {Holland}, {Immler}, {Milne},
  {Roming}, {Gehrels}, {Nousek}, {Panagia}, {Still}, and {Vanden
  Berk}]{brown2009ultraviolet}
{Brown}, P.J.; {Holland}, S.T.; {Immler}, S.; {Milne}, P.; {Roming}, P.W.A.;
  {Gehrels}, N.; {Nousek}, J.; {Panagia}, N.; {Still}, M.; {Vanden Berk}, D.
\newblock Ultraviolet Light Curves of Supernovae with the Swift
  Ultraviolet/Optical Telescope.
\newblock {\em \apj} {\bf 2009}, {\em 137},~4517.

\bibitem[{Tanvir} \em{et~al.}(2017){Tanvir}, {Levan},
  {Gonz{\'a}lez-Fern{\'a}ndez}, {Korobkin}, {Mandel}, {Rosswog}, {Hjorth},
  {D'Avanzo}, {Fruchter}, {Fryer}, and e.a.]{tanvir2017}
{Tanvir}, N.R.; {Levan}, A.J.; {Gonz{\'a}lez-Fern{\'a}ndez}, C.; {Korobkin},
  O.; {Mandel}, I.; {Rosswog}, S.; {Hjorth}, J.; {D'Avanzo}, P.; {Fruchter},
  A.S.; {Fryer}, C.L.; e.a..
\newblock {The Emergence of a Lanthanide-rich Kilonova Following the Merger of
  Two Neutron Stars}.
\newblock {\em \apjl} {\bf 2017}, {\em 848},~L27.

\bibitem[{Levan} \em{et~al.}(2008){Levan}, {Tanvir}, {Jakobsson}, {Chapman},
  {Hjorth}, {Priddey}, {Fynbo}, {Hurley}, {Jensen}, {Johnson}, and
  e.a.]{Levan08GRB050906}
{Levan}, A.J.; {Tanvir}, N.R.; {Jakobsson}, P.; {Chapman}, R.; {Hjorth}, J.;
  {Priddey}, R.S.; {Fynbo}, J.P.U.; {Hurley}, K.; {Jensen}, B.L.; {Johnson},
  R.; e.a..
\newblock {On the nature of the short-duration GRB 050906}.
\newblock {\em \mnras} {\bf 2008}, {\em 384},~541--547.

\bibitem[{Perley} \em{et~al.}(2008){Perley}, {Bloom}, {Modjaz}, {Miller},
  {Shiode}, {Brewer}, {Starr}, and {Kennedy}]{070809GCN}
{Perley}, D.A.; {Bloom}, J.S.; {Modjaz}, M.; {Miller}, A.A.; {Shiode}, J.;
  {Brewer}, J.; {Starr}, D.; {Kennedy}, R.
\newblock {GRB 070809: putative host galaxy and redshift.}
\newblock {\em GRB Coordinates Network} {\bf 2008}, {\em 7889}.

\bibitem[{O'Brien} and {Tanvir}(2009)]{090417GCN}
{O'Brien}, P.T.; {Tanvir}, N.R.
\newblock {GRB 090417A: nearby galaxy redshift.}
\newblock {\em GRB Coordinates Network} {\bf 2009}, {\em 9136}.

\bibitem[{Huchra} \em{et~al.}(2012){Huchra}, {Macri}, {Masters}, {Jarrett},
  {Berlind}, {Calkins}, {Crook}, {Cutri}, {Erdo\u{g}du}, {Falco}, and
  e.a.]{Huchra2012}
{Huchra}, J.P.; {Macri}, L.M.; {Masters}, K.L.; {Jarrett}, T.H.; {Berlind}, P.;
  {Calkins}, M.; {Crook}, A.C.; {Cutri}, R.; {Erdo\u{g}du}, P.; {Falco}, E.;
  e.a..
\newblock {The 2MASS Redshift Survey -- Description and Data Release}.
\newblock {\em \apjs} {\bf 2012}, {\em 199},~26.

\bibitem[{Makarov} \em{et~al.}(2014){Makarov}, {Prugniel}, {Terekhova},
  {Courtois}, and {Vauglin}]{HyperLEDAIII}
{Makarov}, D.; {Prugniel}, P.; {Terekhova}, N.; {Courtois}, H.; {Vauglin}, I.
\newblock {HyperLEDA. III. The catalogue of extragalactic distances}.
\newblock {\em \aap} {\bf 2014}, {\em 570},~A13.

\bibitem[Helou \em{et~al.}(1991)Helou, Madore, Schmitz, Bicay, Wu, and
  Bennett]{helou1991nasa}
Helou, G.; Madore, B.F.; Schmitz, M.; Bicay, M.D.; Wu, X.; Bennett, J.
\newblock The NASA/IPAC extragalactic database. In {\em Databases \& On-Line
  Data in Astronomy}; Springer,  1991; pp. 89--106.

\bibitem[{Tully} \em{et~al.}(2016){Tully}, {Courtois}, and
  {Sorce}]{Cosmicflows3}
{Tully}, R.B.; {Courtois}, H.M.; {Sorce}, J.G.
\newblock {Cosmicflows-3}.
\newblock {\em \aj} {\bf 2016}, {\em 152},~50.

\bibitem[{Fong} \em{et~al.}(2013){Fong}, {Berger}, {Chornock}, {Margutti},
  {Levan}, {Tanvir}, {Tunnicliffe}, {Czekala}, {Fox}, {Perley}, and
  e.a.]{Fong2013}
{Fong}, W.; {Berger}, E.; {Chornock}, R.; {Margutti}, R.; {Levan}, A.J.;
  {Tanvir}, N.R.; {Tunnicliffe}, R.L.; {Czekala}, I.; {Fox}, D.B.; {Perley},
  D.A.; e.a..
\newblock {Demographics of the Galaxies Hosting Short-duration Gamma-Ray
  Bursts}.
\newblock {\em \apj} {\bf 2013}, {\em 769},~56.

\bibitem[{Tanvir} and {Malesani}(2011)]{Tanvir2011GCN}
{Tanvir}, N.R.; {Malesani}, D.
\newblock {GRB 111210A: SDSS prior imaging.}
\newblock {\em GRB Coordinates Network, Circular Service, No.~12661, \#1
  (2011)} {\bf 2011}, {\em 12661}.

\bibitem[{Levan} and {Tanvir}(2013)]{Levan2013GCN}
{Levan}, A.J.; {Tanvir}, N.R.
\newblock {GRB 130515A: FORS2 spectroscopy of candidate counterpart.}
\newblock {\em GRB Coordinates Network, Circular Service, No.~14667, \#1
  (2013)} {\bf 2013}, {\em 14667}.

\bibitem[{Perley} \em{et~al.}(2008){Perley}, {Bloom}, {Modjaz}, {Miller},
  {Shiode}, {Brewer}, {Starr}, and {Kennedy}]{Perley2008GCN}
{Perley}, D.A.; {Bloom}, J.S.; {Modjaz}, M.; {Miller}, A.A.; {Shiode}, J.;
  {Brewer}, J.; {Starr}, D.; {Kennedy}, R.
\newblock {GRB 070809: putative host galaxy and redshift.}
\newblock {\em GRB Coordinates Network} {\bf 2008}, {\em 7889}.

\bibitem[{Fong} \em{et~al.}(2012){Fong}, {Berger}, {Margutti}, {Zauderer},
  {Troja}, {Czekala}, {Chornock}, {Gehrels}, {Sakamoto}, {Fox}, and
  e.a.]{Fong2012GRB111020A}
{Fong}, W.; {Berger}, E.; {Margutti}, R.; {Zauderer}, B.A.; {Troja}, E.;
  {Czekala}, I.; {Chornock}, R.; {Gehrels}, N.; {Sakamoto}, T.; {Fox}, D.B.;
  e.a..
\newblock {A Jet Break in the X-Ray Light Curve of Short GRB 111020A:
  Implications for Energetics and Rates}.
\newblock {\em \apj} {\bf 2012}, {\em 756},~189.

\bibitem[{Cenko} and {Cucchiara}(2013)]{Cenko2013GCN}
{Cenko}, S.B.; {Cucchiara}, A.
\newblock {GRB 130515A: further Gemini observations.}
\newblock {\em GRB Coordinates Network, Circular Service, No.~14670, \#1
  (2013)} {\bf 2013}, {\em 14670}.

\bibitem[{Connaughton} \em{et~al.}(2015){Connaughton}, {Briggs}, {Goldstein},
  {Meegan}, {Paciesas}, {Preece}, {Wilson-Hodge}, {Gibby}, {Greiner}, {Gruber},
  {Jenke}, {Kippen}, and e.a.]{connaughton2015}
{Connaughton}, V.; {Briggs}, M.S.; {Goldstein}, A.; {Meegan}, C.A.; {Paciesas},
  W.S.; {Preece}, R.D.; {Wilson-Hodge}, C.A.; {Gibby}, M.H.; {Greiner}, J.;
  {Gruber}, D.; {Jenke}, P.; {Kippen}, R.M.; e.a..
\newblock {Localization of Gamma-Ray Bursts Using the Fermi Gamma-Ray Burst
  Monitor}.
\newblock {\em The Astrophysical Journal Supplement Series} {\bf 2015}, {\em
  216},~32.

\bibitem[{Briggs} \em{et~al.}(1999){Briggs}, {Pendleton}, {Kippen}, {Brainerd},
  {Hurley}, {Connaughton}, and {Meegan}]{briggs1999error}
{Briggs}, M.S.; {Pendleton}, G.N.; {Kippen}, R.M.; {Brainerd}, J.J.; {Hurley},
  K.; {Connaughton}, V.; {Meegan}, C.A.
\newblock The error distribution of BATSE gamma-ray burst locations.
\newblock {\em \apjs} {\bf 1999}, {\em 122},~503.

\bibitem[{Karachentsev} \em{et~al.}(2013){Karachentsev}, {Makarov}, and
  {Kaisina}]{karachentsev2013}
{Karachentsev}, I.D.; {Makarov}, D.I.; {Kaisina}, E.I.
\newblock {Updated Nearby Galaxy Catalog}.
\newblock {\em \aj} {\bf 2013}, {\em 145},~101.

\bibitem[Lazzati \em{et~al.}(2005)Lazzati, Ghirlanda, and
  Ghisellini]{lazzati2005soft}
Lazzati, D.; Ghirlanda, G.; Ghisellini, G.
\newblock Soft gamma-ray repeater giant flares in the BATSE short gamma-ray
  burst catalogue: constraints from spectroscopy.
\newblock {\em \mnras} {\bf 2005}, {\em 362},~L8--L12.

\bibitem[{Popov} and {Stern}(2006)]{popov2006}
{Popov}, S.B.; {Stern}, B.E.
\newblock {Soft gamma repeaters outside the Local Group}.
\newblock {\em \mnras} {\bf 2006}, {\em 365},~885--890.

\bibitem[{Nakar} \em{et~al.}(2006){Nakar}, {Gal-Yam}, {Piran}, and
  {Fox}]{nakar2006}
{Nakar}, E.; {Gal-Yam}, A.; {Piran}, T.; {Fox}, D.B.
\newblock {The Distances of Short-Hard Gamma-Ray Bursts and the Soft Gamma-Ray
  Repeater Connection}.
\newblock {\em \apj} {\bf 2006}, {\em 640},~849--853.

\bibitem[{Ofek}(2007)]{ofek2007}
{Ofek}, E.O.
\newblock {Soft Gamma-Ray Repeaters in Nearby Galaxies: Rate, Luminosity
  Function, and Fraction among Short Gamma-Ray Bursts}.
\newblock {\em \apj} {\bf 2007}, {\em 659},~339--346.

\bibitem[{Tikhomirova} \em{et~al.}(2010){Tikhomirova}, {Pozanenko}, and
  {Hurley}]{tikhomirova2010}
{Tikhomirova}, Y.Y.; {Pozanenko}, A.S.; {Hurley}, K.S.
\newblock {Search for nearby host galaxies of short gamma-ray bursts detected
  and well localized by BATSE/IPN}.
\newblock {\em Astronomy Letters} {\bf 2010}, {\em 36},~231--236.

\bibitem[{Svinkin} \em{et~al.}(2015){Svinkin}, {Hurley}, {Aptekar},
  {Golenetskii}, and {Frederiks}]{svinkin2015search}
{Svinkin}, D.S.; {Hurley}, K.; {Aptekar}, R.L.; {Golenetskii}, S.V.;
  {Frederiks}, D.D.
\newblock {A search for giant flares from soft gamma-ray repeaters in nearby
  galaxies in the Konus-WIND short burst sample}.
\newblock {\em \mnras} {\bf 2015}, {\em 447},~1028--1032.

\bibitem[{Evans} \em{et~al.}(1980){Evans}, {Klebesadel}, {Laros}, {Cline},
  {Desai}, {Teegarden}, {Pizzichini}, {Hurley}, {Niel}, and
  {Vedrenne}]{evans1980location}
{Evans}, W.D.; {Klebesadel}, R.W.; {Laros}, J.G.; {Cline}, T.L.; {Desai}, U.D.;
  {Teegarden}, B.J.; {Pizzichini}, G.; {Hurley}, K.; {Niel}, M.; {Vedrenne}, G.
\newblock Location of the gamma-ray transient event of 1979 March 5.
\newblock {\em \apj} {\bf 1980}, {\em 237},~L7--L9.

\bibitem[{Cline} \em{et~al.}(1981){Cline}, {Desai}, {Teegarden}, {Evans},
  {Klebesadel}, {Laros}, {Barat}, {Hurley}, {Niel}, {Bedrenne}, and
  e.a.]{cline1981precise}
{Cline}, T.L.; {Desai}, U.D.; {Teegarden}, B.J.; {Evans}, W.D.; {Klebesadel},
  R.W.; {Laros}, J.G.; {Barat}, C.; {Hurley}, K.; {Niel}, M.; {Bedrenne}, G.;
  e.a..
\newblock Precise source location of the anomalous 1979 March 5 gamma ray
  transient {\bf 1981}.

\bibitem[{Perley} and {Bloom}(2007)]{Perley2007gcn}
{Perley}, D.A.; {Bloom}, J.S.
\newblock {GRB 070201: proximity of IPN annulus to M31.}
\newblock {\em GRB Coordinates Network} {\bf 2007}, {\em 6091}.

\bibitem[{Hurley} \em{et~al.}(2010){Hurley}, {Rowlinson}, {Bellm}, {Perley},
  {Mitrofanov}, {Golovin}, {Kozyrev}, {Litvak}, {Sanin}, {Boynton}, and
  e.a.]{hurley2010new}
{Hurley}, K.; {Rowlinson}, A.; {Bellm}, E.; {Perley}, D.; {Mitrofanov}, I.G.;
  {Golovin}, D.V.; {Kozyrev}, A.S.; {Litvak}, M.L.; {Sanin}, A.B.; {Boynton},
  W.; e.a..
\newblock A new analysis of the short-duration, hard-spectrum GRB 051103, a
  possible extragalactic soft gamma repeater giant flare.
\newblock {\em \mnras} {\bf 2010}, {\em 403},~342--352.

\bibitem[{Levan} \em{et~al.}(2015){Levan}, {Tanvir}, and
  {Hjorth}]{Levan2015gcn}
{Levan}, A.J.; {Tanvir}, N.R.; {Hjorth}, J.
\newblock {Short GRB 150906B: proximity to NGC 3313 galaxy group.}
\newblock {\em GRB Coordinates Network, Circular Service, No.~18263, \#1
  (2015)} {\bf 2015}, {\em 18263}.

\bibitem[{Abbott} \em{et~al.}(2008){Abbott}, {Abbott}, {Adhikari}, {Agresti},
  {Ajith}, {Allen}, {Amin}, {Anderson}, {Anderson}, {Arain}, and
  e.a.]{abbott2008implications}
{Abbott}, B.; {Abbott}, R.; {Adhikari}, R.; {Agresti}, J.; {Ajith}, P.;
  {Allen}, B.; {Amin}, R.; {Anderson}, S.B.; {Anderson}, W.G.; {Arain}, M.;
  e.a..
\newblock Implications for the Origin of GRB 070201 from LIGO Observations.
\newblock {\em \apj} {\bf 2008}, {\em 681},~1419.

\bibitem[{Abadie} \em{et~al.}(2012){Abadie}, {Abbott}, and
  {Abbott}]{abadie2012}
{Abadie}, J.; {Abbott}, B.P.; {Abbott}, T.D.e.
\newblock {Implications for the Origin of GRB 051103 from LIGO Observations}.
\newblock {\em \apj} {\bf 2012}, {\em 755},~2.

\bibitem[{Abbott} \em{et~al.}(2017{\natexlab{a}}){Abbott}, {Abbott}, {Abbott},
  {Abernathy}, {Acernese}, {Ackley}, {Adams}, {Adams}, {Addesso}, {Adhikari},
  and e.a.]{abbott2017search}
{Abbott}, B.P.; {Abbott}, R.; {Abbott}, T.D.; {Abernathy}, M.R.; {Acernese},
  F.; {Ackley}, K.; {Adams}, C.; {Adams}, T.; {Addesso}, P.; {Adhikari}, R.X.;
  e.a..
\newblock Search for gravitational waves associated with gamma-ray bursts
  during the first advanced LIGO observing run and implications for the origin
  of GRB 150906B.
\newblock {\em \apj} {\bf 2017}, {\em 841},~89.

\bibitem[{Abbott} \em{et~al.}(2017{\natexlab{b}}){Abbott}, {Abbott}, {Abbott},
  {Acernese}, {Ackley}, {Adams}, {Adams}, {Addesso}, {Adhikari}, {Adya}, and
  e.a.]{abbott2017gw170817}
{Abbott}, B.P.; {Abbott}, R.; {Abbott}, T.D.; {Acernese}, F.; {Ackley}, K.;
  {Adams}, C.; {Adams}, T.; {Addesso}, P.; {Adhikari}, R.X.; {Adya}, V.B.;
  e.a..
\newblock GW170817: Observation of Gravitational Waves from a Binary Neutron
  Star Inspiral.
\newblock {\em Physical Review Letters} {\bf 2017}, {\em 119},~161101.

\bibitem[{Chruslinska} \em{et~al.}(2018){Chruslinska}, {Belczynski}, {Klencki},
  and {Benacquista}]{Chruslinska2018}
{Chruslinska}, M.; {Belczynski}, K.; {Klencki}, J.; {Benacquista}, M.
\newblock {Double neutron stars: merger rates revisited}.
\newblock {\em \mnras} {\bf 2018}, {\em 474},~2937--2958.

\bibitem[{Troja} \em{et~al.}(2017){Troja}, {Lipunov}, {Mundell}, {Butler},
  {Watson}, {Kobayashi}, {Cenko}, {Marshall}, {Ricci}, {Fruchter}, and
  e.a.]{troja2017}
{Troja}, E.; {Lipunov}, V.M.; {Mundell}, C.G.; {Butler}, N.R.; {Watson}, A.M.;
  {Kobayashi}, S.; {Cenko}, S.B.; {Marshall}, F.E.; {Ricci}, R.; {Fruchter},
  A.; e.a..
\newblock {Significant and variable linear polarization during the prompt
  optical flash of GRB 160625B.}
\newblock {\em \nat} {\bf 2017}, {\em 547},~425--427.

\bibitem[{Hallinan} \em{et~al.}(2017){Hallinan}, {Corsi}, {Mooley},
  {Hotokezaka}, {Nakar}, {Kasliwal}, {Kaplan}, {Frail}, {Myers}, {Murphy}, and
  e.a.]{hallinan2017}
{Hallinan}, G.; {Corsi}, A.; {Mooley}, K.P.; {Hotokezaka}, K.; {Nakar}, E.;
  {Kasliwal}, M.M.; {Kaplan}, D.L.; {Frail}, D.A.; {Myers}, S.T.; {Murphy}, T.;
  e.a..
\newblock {A radio counterpart to a neutron star merger}.
\newblock {\em Science} {\bf 2017}, {\em 358},~1579--1583.

\bibitem[{Lyman} \em{et~al.}(2018){Lyman}, {Lamb}, {Levan}, {Mandel}, {Tanvir},
  {Kobayashi}, {Gompertz}, {Hjorth}, {Fruchter}, {Kangas}, and e.a.]{lyman2018}
{Lyman}, J.D.; {Lamb}, G.P.; {Levan}, A.J.; {Mandel}, I.; {Tanvir}, N.R.;
  {Kobayashi}, S.; {Gompertz}, B.; {Hjorth}, J.; {Fruchter}, A.S.; {Kangas},
  T.; e.a..
\newblock {The optical afterglow of the short gamma-ray burst associated with
  GW170817}.
\newblock {\em Nature Astronomy} {\bf 2018}, {\em 2},~751--754.

\bibitem[{Evans} \em{et~al.}(2017){Evans}, {Cenko}, {Kennea}, {Emery}, {Kuin},
  {Korobkin}, {Wollaeger}, {Fryer}, {Madsen}, {Harrison}, and e.a.]{evans2017}
{Evans}, P.A.; {Cenko}, S.B.; {Kennea}, J.A.; {Emery}, S.W.K.; {Kuin}, N.P.M.;
  {Korobkin}, O.; {Wollaeger}, R.T.; {Fryer}, C.L.; {Madsen}, K.K.; {Harrison},
  F.A.; e.a..
\newblock {Swift and NuSTAR observations of GW170817: Detection of a blue
  kilonova}.
\newblock {\em Science} {\bf 2017}, {\em 358},~1565--1570.

\bibitem[{Licquia} and {Newman}(2015)]{licquia2015}
{Licquia}, T.C.; {Newman}, J.A.
\newblock {Improved Estimates of the Milky Way's Stellar Mass and Star
  Formation Rate from Hierarchical Bayesian Meta-Analysis}.
\newblock {\em \apj} {\bf 2015}, {\em 806},~96.

\bibitem[{Abbott} \em{et~al.}(2018){Abbott}, {Abbott}, {Abbott}, {Abernathy},
  {Acernese}, {Ackley}, {Adams}, {Adams}, {Addesso}, {Adhikari}, and
  e.a.]{abbottLRR}
{Abbott}, B.P.; {Abbott}, R.; {Abbott}, T.D.; {Abernathy}, M.R.; {Acernese},
  F.; {Ackley}, K.; {Adams}, C.; {Adams}, T.; {Addesso}, P.; {Adhikari}, R.X.;
  e.a..
\newblock {Prospects for observing and localizing gravitational-wave transients
  with Advanced LIGO, Advanced Virgo and KAGRA}.
\newblock {\em Living Reviews in Relativity} {\bf 2018}, {\em 21},~3.

\end{thebibliography}
\end{document}